\documentclass[journal]{IEEEtran}
\usepackage{comment}
\usepackage{amsmath,amssymb,bm}
\usepackage{tabularx}
\usepackage[english]{babel}

\usepackage{graphicx}
\usepackage{booktabs} 
\usepackage{caption}
\usepackage{subcaption}
\usepackage{xcolor}
\usepackage{cite}
\usepackage{bbm}
\usepackage{algorithmic}
\usepackage[linesnumbered,ruled]{algorithm2e}
\usepackage[colorlinks=true,  linkcolor=black, citecolor=blue, breaklinks=true, urlcolor=blue]{hyperref}

\newcommand{\tran}{^{\mbox{\scriptsize T}}}
\newcommand{\herm}{^{\mbox{\scriptsize H}}}

\usepackage{tabularx}
\usepackage{pgfplots}
\usepackage{tikz}
\usepgfplotslibrary{groupplots}
\pgfplotsset{compat=1.13}
\newlength\fwidth

\newcommand{\C}{\mathbb{C}} 
\renewcommand{\b}{{\bo b}}

\newcommand{\bo}[1]{\bm{#1}}              
\newcounter{cremark}
\newtheorem{remark}[cremark]{Remark}
\newcommand{\SCM}{\hat{\boldsymbol{\Sigma}}}
\newcommand{\Gam}{\boldsymbol{\Gamma}}

\newcommand{\B}{\bm{B}}
\newcommand{\supp}{\mathsf{supp}}

\renewcommand{\dim}{L}  
\newcommand{\ndata}{M}  
\newcommand{\natoms}{N} 

\newcommand{\hop}{\mathsf{H}}        

\newcommand{\beq}{\begin{equation}}
\newcommand{\eeq}{\end{equation}}
\newcommand{\bmat}{\begin{pmatrix}}
\newcommand{\emat}{\end{pmatrix}}
\newcommand{\gam}{\boldsymbol{\gamma}}

\newcommand{\X}{\bm{X}}
\newcommand{\A}{\bm{A}}
\newcommand{\Y}{\bm{Y}}

\newcommand{\W}{\bm{W}}

\renewcommand{\S}{\hat{\boldsymbol{\Sigma}}} 

\newcommand{\support}{\mathcal{K}}

\newcommand{\y}{\bo y}
\newcommand{\x}{\bm{x}}

\renewcommand{\a}{{\bo a}}

\newcommand{\M}{\boldsymbol{\Sigma}}

\DeclareMathOperator{\tr}{tr}

\DeclareMathOperator{\diag}{diag}

\newcommand{\nonl}{\renewcommand{\nl}{\let\nl\oldnl}}

\ifCLASSINFOpdf
\else
\fi
\begin{document}
\title{Activity Detection for Massive Random Access using Covariance-based Matching Pursuit}
\author{\IEEEauthorblockN{Leatile Marata, \textit{Member, IEEE}, Esa Ollila, \textit{Senior Member, IEEE}, and Hirley Alves, \textit{Member, IEEE}}
\thanks{Leatile Marata, and Hirley Alves are with Centre for Wireless Communications -- Radio Technologies, FI-90014, University of Oulu, Finland. e-mail: \{leatile.marata, hirley.alves\}@oulu.fi.  Esa Ollila is with the Department of Information and Communications Engineering, Aalto University, FI-00076 Aalto, Finland (e-mail: esa.ollila@aalto.fi)}
\thanks{This work is supported by the Research Council of Finland (former Academy of Finland) (Grants n.319485, n.340171, n.346208 (6G Flagship), n.338408, n.353093, n.359848). Leatile Marata's work was partly supported by the Riitta ja Jorma J. Takanen Foundation, the Nokia Foundation, the Finnish Foundation for Technology Promotion, and the Botswana International University of Science and Technology.}}
\maketitle
\begin{abstract}
    The Internet of Things paradigm heavily relies on a network of a massive number of machine-type devices (MTDs) that monitor various phenomena. Consequently, MTDs are randomly activated at different times whenever a change occurs. In general, fewer MTDs are simultaneously activated across the network, resembling targeted sampling in compressed sensing. Therefore, signal recovery in machine-type communications is addressed through joint user activity detection and channel estimation algorithms built using compressed sensing theory. However, most of these algorithms follow a two-stage procedure in which a channel is first estimated and later mapped to find active users. This approach is inefficient because the estimated channel information is subsequently discarded. To overcome this limitation, we introduce a novel covariance-learning matching pursuit (CL-MP) algorithm that bypasses explicit channel estimation. Instead, it focuses on estimating the indices of the active users greedily. Simulation results presented in terms of probability of miss detection, exact recovery rate, and computational complexity validate the proposed technique's superior performance and efficiency.
\end{abstract}

\begin{IEEEkeywords} Activity detection, compressed sensing, covariance-learning, grant-free, NOMA, matching pursuit, sporadic activation. 
\end{IEEEkeywords}
\IEEEpeerreviewmaketitle

\section{Introduction}
\label{Introduction}
\IEEEPARstart{M}{a}ssive machine-type communications (mMTC) is a framework that enables nonhuman devices to exchange information with each other without requiring human intervention \cite{shariatmadari2015machine,dutkiewicz2017massive,marata2022joint}. This kind of communication is predominantly uplink-driven and is characterized by devices transmitting their data to a central point such as a base station (BS) in the cellular Internet of Things (IoT). Indeed, this is particularly crucial for the  IoT, where many machine-type devices (MTDs), e.g., sensors and actuators
work hand in hand to collect measurements of different phenomena and perform automatic tasks \cite{shariatmadari2015machine,marata2023joint}. For instance, the sixth generation (6G) wireless communications standard is expected to host a minimum ratio of $10^8$ MTDs per square kilometer. This necessitates the development of novel efficient signal processing techniques and adaption of existing ones to work in mMTC \cite{gao2023energy}. Given the massive communication scale, next-generation multiple access techniques are fundamental in achieving ubiquitous connectivity of mMTC \cite{liu2022evolution}. Such techniques include the grant-free (GF) non-orthogonal multiple access (NOMA) techniques, which have been widely adopted as a key enabler for MTC \cite{shariatmadari2015machine,dutkiewicz2017massive}. These schemes reduce network access latency by bypassing the handshaking process that is normally used in grant-based systems to coordinate access to the communication medium \cite{choi2021grant}. Unfortunately, such latency reduction comes at the expense of an increased number of transmitted data collisions, which are ultimately detrimental to the performance of BS receivers \cite{senel2018grant}. In the end, the efficiency of mMTC (and consequently IoT) networks is subject to how well the collisions in media access can be resolved. To address this, Liu \textit{et al.} in \cite{liu2018sparse} proposed the compressed sensing (CS)- multi-user detection (MUD) framework for MTC receivers. 
For instance, the CS-MUD frameworks in  \cite{li2021compressed, liu2020modeling,zhang2022bayesian}  follow a two-stage procedure where the sparse channel matrix is first estimated, followed by a thresholding step. 
The thresholding step estimates the activity indicator indices by assigning a one to the rows of the channel matrix estimate whose norms exceed the threshold and a zero to those below it, thereby determining whether an MTD is active or inactive.
Among the most widely adopted solutions are the variants of approximate message passing (AMP), sparse Bayesian learning  (SBL), and simultaneous orthogonal matching pursuit (SOMP) \cite{di2020detection}.

\par Although the two-stage signal recovery approaches mentioned above are applicable in many cases, they become inefficient when the primary goal is to identify active devices rather than estimating the channels. 
For example, applications like industrial process monitoring often rely on collective decision-making based on averaged measurements. In such cases, estimating indices of hardwired codebook entries for each MTD is typically sufficient, thereby eliminating the need for channel estimation \cite{polyanskiy2017perspective,shao2020cooperative,kalor2024wireless}. Equally important is the fact that the counterpart of the two-stage signal recovery,  called non-coherent detection, makes it necessary to estimate the indices explicitly \cite{ni2020non}. Furthermore, as mMTC deployments continue to expand, the overarching challenge extends beyond merely resolving collisions to developing algorithms that are both effective and efficient \cite{lopez2023statistical}. As such, bypassing explicit channel estimation can facilitate the development of faster algorithms that use less computational resources. To this end, our work focuses on developing a novel receiver algorithm that enables optimal active user identification by estimating the active user indices in MTC. To provide context, we summarise the state-of-the-art works relevant to our study in the next sub-section. 
{\subsection{State of the Art}
\label{stateofArtFin}
Research on the problem of active user detection (AUD) and channel estimation for MTC has been gaining popularity due to its importance in enabling IoT. There are various approaches to addressing this problem. For instance, the Bayesian formulation, which defines probability distributions for variables of interest, such as sparsity-inducing parameters, has been explored in \cite{di2020detection,liu2018sparse}. Notably, Zhang \textit{et al.} proposed an AMP algorithm combined with variational Bayesian inference for joint user detection and data recovery in \cite{zhang2023variational}. It is worth mentioning that while algorithms developed based on AMP are less computationally demanding, they are highly sensitive to non-Gaussian sensing matrices, which limits their practical implementation. On the other hand, algorithms  based on SBL are computationally expensive but more robust to the design of the sensing matrix. To leverage this robustness with less computational resources, Zhang \textit{et al.} proposed a fast inversion SBL for GF-NOMA for MTC in \cite{zhang2021joint}, aimed at lowering the computational complexity of classical SBL. On the other hand, Wang \textit{et al.} proposed a variational Bayesian inference algorithm in \cite{wang2020variational} for joint user activity detection in mMTC that is supported by a cloud-based radio access network. 

\par Another approach to solving the AUD problem follows the convex optimization framework, utilizing the $\ell_1$-norm minimization. Notable works in this direction include \cite{djelouat2021joint}, in which Djelouat \textit{et al.} proposed the alternating direction method of multipliers (ADMM) for device activity detection in MTC with correlated channels. Sun \textit{et al.} also proposed a joint user activity detection optimization algorithm in \cite{sun2023joint} by exploiting the joint sparsity of the sensing matrix and the effective channels of the MTDs. Similarly, Zhu \textit{et al.} introduced a joint user activity detection method in \cite{zhu2023active}, employing convex optimization based on atomic norm minimization. 

In addition to the aforementioned approaches, the covariance-based activity detection approach has also gained popularity due to its appealing scalability \cite{chen2019covariance,shao2020cooperative,liu2024grant}. For instance, Liu \textit{et al.} proposed a maximum likelihood estimation (MLE) approach in \cite{liu2024mle} for device activity detection in channels with line of sight (LoS). The proposed solution is shown to reduce the probability of error by $50\%$ with reduced computation times. On the other hand, Fengler \textit{et al.} proposed a low complexity MLE algorithm in \cite{fengler2021non}, and derived a  scaling law for the number of identifiable active users.  Meanwhile, You \textit{et al.} proposed a covariance-based algorithm for unsourced random access in \cite{you2023efficient}. Similarly, Yu \textit{et al.} in \cite{yu2024joint} proposed sufficient conditions for solving the maximum likelihood problem using a covariance-based approach. 

\par Despite the solutions discussed above, most of them first estimate the effective channels induced by sporadic activation and rather than explicitly estimating the active users, with the exceptions of \cite{liu2024mle,shao2020cooperative}. Unfortunately, the works in \cite{liu2024mle,shao2020cooperative} face inherent limitations due to their reliance on  cyclic coordinate descent minimization, resulting in an iterative algorithm that can suffer from slow convergence. Moreover, most existing algorithms, except for SBL, require some form of parameter tuning and are not robust to some pilot sequences, limiting their applicability. As mentioned earlier, the drawback of SBL is its prohibitively high computational complexity, making it unsuitable for massive MTC (mMTC).  Therefore, we aim to propose a novel solution for receivers in mMTC using a greedy search approach. Unlike coordinate descent approaches, greedy search is not subject to convergence issues, and is generally faster to compute than optimization-based approaches. 

The proposed algorithm,  called covariance-learning matching pursuit (CL-MP),  is a fast and enhanced modification of the CL-OMP algorithm developed recently in \cite{ollila2024sparse}. It is important to highlight that the CL-MP differs from \cite{ollila2024sparse} in several ways and incorporates features that make it better suited for the AUD problem in mMTC. These differences are discussed in more detail in Remark~\ref{rem:CLMPcomparison} in \autoref{sec:MP}.  
Ultimately, the proposed method detects active devices with low computational complexity and does not require parameter tuning. 

\subsection{Contributions}
\label{contributions}
We consider an mMTC scenario where sporadically activated MTDs transmit their data in the uplink and propose a covariance-learning (CL) matching pursuit (MP) algorithm, termed CL-MP, to recover the indices of active devices. The work departs from 
classical covariance-based techniques that rely on coordinate-wise optimization (CWO) algorithms such as \cite{chen2019covariance,fengler2021non}. Our main contributions are summarized as follows:  
\begin{itemize}
    \item We formulate the device activity detection problem as the minimization of the Gaussian negative log-likelihood function and solve it using the CL-MP algorithm. It is worth noting that 
    SBL, CWO, and CL-MP have similar cost functions. 
    Consequently, the CL-MP algorithm offers the same performance in terms of user identification as SBL and CWO. However, CL-MP is more efficient compared to SBL and CWO. 

     \item We evaluate the CL-MP algorithm using various pilot sequences, including structured pilot sequences proposed in \cite{marata2024advanced,marata2023joint}, as well as Bernoulli and Gaussian pilot sequences. While the performance may vary across different pilot sequences, CL-MP consistently yields reasonable results under these different pilot sequences. 
\item Finally, 
We compare our proposed solution with existing techniques, such as SOMP, SBL, and VAMP. The results reveal that CL-MP outperforms SOMP and VAMP while achieving performance comparable to SBL in terms of detection capabilities. However, the runtime results indicate that CL-MP is faster than SBL.
\end{itemize}

\subsection{Organization and Notation}

The remainder of this paper is organized as follows: Section~\ref{formulate} introduces the system model and the AUD problem. In Section~\ref{sec:MP}, we present the solution to the device activity detection problem using CL-MP. Lastly, in Section~\ref{Results}, we present the results and discussions, while in Section~\ref{conclusionAndFuture}, we summarize the paper and outline potential future research directions. 

\par 
\textit{\textbf{Notation:}} Boldface lowercase and boldface uppercase letters denote column vectors and matrices, respectively. Moreover, for a matrix $\A$, $\a_i$ is the $i^{\text{th}}$ column, while $a_{ij}$ denotes the element in the $i^{\text{th}}$ row, $j^{\text{th}}$  column, respectively, whereas $a_i$ denotes the $i^{\text{th}}$ element of a vector $\a$. 
We denote the transpose and conjugate transpose operations by superscripts 
$(\cdot)\tran$ and $(\cdot)\herm$, respectively, while $\mathbb{C}$ and  $ \mathbb{R}$ refer to complex and real domains, respectively. 
The circularly symmetric $M$-variate complex Gaussian distribution with mean $\a$ and covariance matrix $\M$ is denoted by $\mathcal{CN}_{\!M}(\a,\M)$. Let  $\diag(\bm{a})$ denote a diagonal matrix whose diagonal elements are vector $\a$. Meanwhile
 $\bm{I}$ is the identity matrix, while  
$(\cdot)_{+} = \max(\cdot,0)$ is a non-negativity function,  and $[\! [  M  ] \!]  = \{1,\ldots, M \}$.

\section{SYSTEM MODEL}\label{formulate}

\begin{figure}
    \centering    \includegraphics[scale= 0.5]{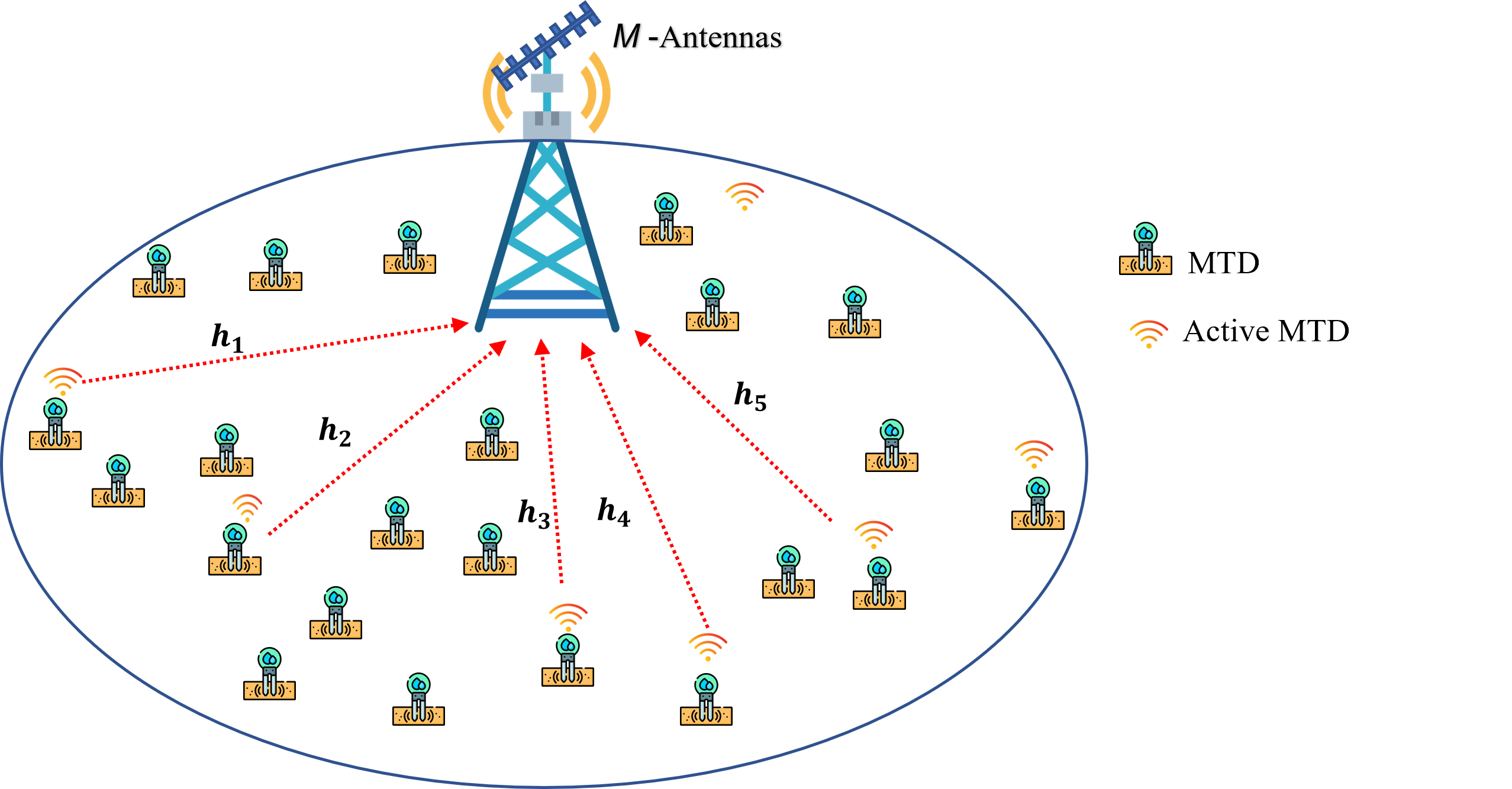}
    \caption{A massive machine-type communications network scenario, where $N$ MTDs transmit in the uplink to a BS equipped with $M$ antennas.}
    \label{systemModelFig}
\end{figure} 
We consider a communication network scenario depicted by \autoref{systemModelFig}, where a BS equipped with  
$M$ antennas serves a set of $N$
MTDs that transmit data in the uplink. To align with practical mMTC, we assume that in a given coherence interval of $T$ symbols, an unknown set $\support \subset [\![ N]\!]$ of cardinality $K = |\mathcal{K}|$ is active. 
Note that from these $T$ symbols of the coherence interval, $L$ symbols are used for metadata processing to identify the $K$ active MTDs. Meanwhile, the remaining $T-L$ symbols are used to transmit their payload data. Furthermore, we assume that each $n^{\text{th}}$ MTD is connected to the BS through a quasi-static fading uplink channel $\bm{h}_n \in \mathbb{C}^{M\times 1}$.  For simplicity, we assume an independent Rayleigh small-scale fading model so that the channel vectors are statistically independent
of each other and spatially white, i.e., $\bm{h}_{n} \sim \mathcal{CN}_{\!M}(\bm{0},\bm{I})$. 

To account for the random activation of the MTDs, we define the activity indicator function as 
\begin{equation}
  \alpha_{n}\!=\! 
  \begin{cases} 
     1,~\text{if device $n$ is active}\\     0,~\text{otherwise}\\
   \end{cases}.  
   \label{actFuction}
\end{equation}
As a result of \eqref{actFuction}, $\alpha_{n}$ induces an effective channel $\x_{n} = \sqrt{\gamma_{n}}\bm{h}_{n}$ for the $n^{\text{th}}$ MTD. Here, $\gamma_n = \alpha_n \rho_n\beta_n$ is the effective power of the channel of the $n$-th MTD, where $\beta_n>0$  is the large-scale fading component (LSFC) of the $n^{\text{th}}$ device due to path-loss and shadowing, while $\rho_n$ is the device's uplink transmission power. Hence, the concatenation of all the effective power levels of the channels of the MTDs in the network is defined by {$\gam=(\gamma_1,\ldots,\gamma_N)^\top\in \mathbb{R}^N$}. Similarly, let  $\X = (\x_{1} \, \cdots \,  \x_{N})\tran \in \mathbb{C}^{N \times M}$ denote the concatenation of all the effective channel vectors. 
Since only a small set of MTDs are active, $\X$ is row-sparse. This leads to the uplink signal being a ``compressively sensed" signal coming exclusively from active devices, as will be seen later. As such, the set of active MTDs corresponds to the row-support of  $\X \in \C^{\natoms \times \ndata}$. The index set  of rows  containing non-zero  elements is defined as
\begin{align*}
\support &= \mathrm{supp}(\X) 
\triangleq \{   i \in  [\! [   \natoms ] \!]   :  x_{nm} \neq 0 \,  \mbox{ for some $m \in [\! [ \ndata ] \!] $} \}, 
\end{align*}
thus collecting the indices of the active devices, i.e., $\support = \{ n \in [\! [   \natoms ] \!]  : \alpha_n =1\}$. Note that, due to the low likelihood of the MTDs being active simultaneously, $\lvert\mathcal{K}\rvert< N$ is a practical network property leveraged in the CS-MUD framework, as will be demonstrated next.

As mentioned earlier, $L$ symbols are used as pilot sequences, i.e., used mainly for metadata processing, such as to estimate the active MTDs. To avoid signaling overheads, these pilot sequences are normally preallocated by the BS and stored in the local memory of the MTDs. Let $\a_n \in \mathbb{C}^{L\times 1}$ denote the normalized pilot sequence of the $n^{\text{th}}$ MTD.
These can be 
collected into a matrix $\A = (\a_1 \, \cdots \,  \a_N)\in \mathbb{C}^{L\times N}$. In the end, the first $L$ symbols of the noisy signal received at the BS, exclusively coming from the $K$ active MTDs, can now be modeled as
\begin{align}
    \Y &= \sum_{k \in \mathcal{K}}\sqrt{\rho_k\beta_{k}}\alpha_{k}\bm{a}_{k}\bm{h}_{k}\tran + \bm{W}\nonumber \\
        &= \A\X + \W,
     \label{signalReceived}
\end{align}
where $\W\in \mathbb{C}^{L \times M}$ is an additive noise matrix whose elements are independent and identically distributed (i.i.d.) Gaussian random variables with known variance $\sigma^2>0$, i.e., $w_{lm}\sim \mathcal{CN}(0,\sigma^2)$. Owing to the massiveness of the MTDs, the length of the pilot sequences is normally very short compared to the number of devices, i.e., $L\ll N$, essentially leading to non-orthogonal pilot sequences that can have noticeable mutual coherence  \cite{marata2022joint,marata2023joint}. Therefore, the task of the BS is to identify the active users by estimating the row-support  $\mathcal{K}$ of $\bm{X}$. However, the condition $L\ll N$ implies that  \eqref{signalReceived} corresponds to 
an underdetermined linear system\footnote{An undetermined system of linear equations is characterized by having more unknowns than the number of equations.} of equations, thus leading to an infinite number of solutions. To overcome this challenge, one must resort to the CS-MUD methodology, such as those discussed in Section~\ref{stateofArtFin}. We propose a greedy pursuit approach for determining the minimizer of the Gaussian negative log-likelihood function, as discussed next. 

\par Since there is no correlation in the noise samples at the receive antennas, the columns $\{ \bm{y}_{m} \in \mathbb{C}^{L}\}_{m =1}^{M}$ of $\bm{Y}$ are independent of each other. This means that $\y_m \sim \mathcal{C N}_{\dim}(\bm{0}, \M)$ has a  positive definite Hermitian (PDH) $\dim \times \dim$ covariance matrix (CM): 
\begin{align}
 \label{eq:M}
\M &= \A \Gam \A^\hop + \sigma^2 \bm{I}\nonumber \\ &=  \sum_{n=1}^{\natoms} \gamma_n \a_n \a_n^\hop + \sigma^2 \bm{I}, 
\end{align}   
where  {$\Gam = \diag(\gam)$} is a diagonal matrix containing the effective powers of the MTDs. Note that   $\support = \mathrm{supp}(\X)=\supp(\gam)$, which makes it possible to estimate the support using \eqref{eq:M}. As such, \emph{covariance learning} (CL-) based support recovery algorithms can be constructed by minimizing the negative log-likelihood function (LLF) of the data $\Y$. The negative LLF is defined (after scaling by $1/\ndata$ and ignoring additive constants) as follows:
\begin{align}  
\ell( \boldsymbol{\gamma}) &=  \tr(  \M^{-1} \SCM )  +  \log | \M |   \label{eq:SMLapu}  \\
&= \tr(  (\A   \Gam\A^\hop + \sigma^2 \bm{I})^{-1} \SCM )  +  \log |  \A \Gam \A^\hop + \sigma^2 \bm{I} |,    \label{eq:SML}
\end{align}
where $\SCM$ is  the sample covariance matrix (SCM),  
\[
\SCM = \frac{1}{\ndata} \sum_{m=1}^{\ndata} \y_{m} \y_m^\hop = {\ndata}^{-1} \bm{Y} \bm{Y}^\hop, 
 \]
while $\tr(\cdot)$ and $| \cdot |$ denote the matrix trace and determinant, respectively.

The LLF in \eqref{eq:SML} resembles the \emph{deterministic} Gaussian LLF model where signals $\x_i$, $i= 1, \ldots, N$ are treated as deterministic yet unknown. 
This implies that each $\y_m$ follows a Gaussian distribution, i.e., $\y_m \sim \mathcal{C N}_{\dim}(\A \x_m, \sigma^2 \mathbf{I})$, such that the  deterministic (scaled) negative Gaussian LLF is 
\beq \label{eq:ellX}
\ell( \X) =  \| \Y - \A \X \|_{\text{F}}^2  , 
\eeq
where $\lVert \cdot \rVert_{F}$ denotes the Frobenius norm. Given the LLF, the objective is to find the row-sparse signal matrix $\X$ that minimize \eqref{eq:ellX}. 
Greedy sparse signal recovery (SSR) methods such as simultaneous matching pursuit  (MP) or simultaneous orthogonal matching pursuit (OMP) aim at solving $\ell(\X)$ subject to the condition that the cardinality of the norms of the rows of $\X$ is $\| \X \|_{2,0} = K$ \cite{duarte2011structured}.  As such, the greedy methods solve the aforementioned optimization problem  by focusing on the support, i.e., greedily selecting the  column of $\A$ at each stage that maximally reduces the residual $L_2$-error when approximating $\Y$ using the currently active columns. In the next section, we develop a greedy approach for the \emph{stochastic} LLF model in \eqref{eq:SML}.  The proposed method is analogous to MP, offering the advantage of being computationally lighter than other similar greedy approaches, such as OMP or Compressive Sampling Matching Pursuit (CoSaMP).

\section{Covariance based matching pursuit} \label{sec:MP} 
Consider the conditional likelihood of \eqref{eq:SML} where device powers  $\gamma_n$ for $n \neq i$ are known, and define 
\beq \label{eq:Mbackslash}
\M_{\backslash i} = \sum_{n \neq i}  \gamma_n   \a_n \a_n^\hop + \sigma^2 \bm{I}  = \M - \gamma_i \a_i \a_i^{\hop} 
\eeq 
as the covariance matrix of  $\y_m$-s when the contribution from the $i^{\text{th}}$ device is removed. As $\gamma_n$ for $n \neq i$ are assumed to be given, $\M_{\backslash i}$ is a known fixed matrix. Thus, under the above assumption, we have $\y_m   \sim \mathcal C \mathcal N_L(\mathbf{0}, \M_{\backslash i} + \gamma  \a_i \a_i^\hop)$, where $\gamma$ is the sole unknown parameter (the $i^{\text{th}}$ device power), and $\M_{\backslash i}$ is known.  Hence, the conditional  negative LLF for the single  unknown $i^{\text{th}}$ device power becomes
\begin{align} \label{eq:epsilon_i_v0_b}
\ell_i(\gamma \mid \M_{\setminus i} ) &=  \tr( (\M_{\setminus i} + \gamma \a_i \a_i^\hop )^{-1}  \S) + \log | \M_{\setminus i} + \gamma \a_i \a_i^\hop |
\end{align}
 which follows by substituting $\M_{\backslash i} + \gamma  \a_i \a_i^\hop$ in place of $\M$ in \eqref{eq:SMLapu}. 
This function has a unique optimal value that is given by   \cite{yardibi2010source}: 
\begin{align} \label{eq:gamma_i_star}
\gamma_i  &=  \arg \min_{\gamma \geq 0}  \ \ell_i(\gamma \mid \M_{\setminus i} ) \notag \\
&=  \max\left( \frac{\a_i^\hop \M_{\setminus i}^{-1}  \S  \M_{\setminus i}^{-1} \a_i - \a_i^\hop \M_{\setminus i}^{-1} \a_i }{(\a_i^\hop \M_{\setminus i}^{-1} \a_i)^2 } , 0\right), 
\end{align} 
where  $\M_{\setminus i}$ is defined in \eqref{eq:Mbackslash}. To make the paper self-contained we provide the proof of this result in 
Appendix~\ref{App:A}.

This result forms the foundation for developing the CL-MP
algorithm. We adopt the generic matching pursuit strategy (see, e.g., \cite{sparse_book:2010}) and  denote the $k^{\text{th}}$ update of $\boldsymbol{\gamma}$ and $\mathcal{K}$ as $\boldsymbol{\gamma}^{(k)}$ and $\mathcal{K}^{(k)}$, respectively.
Starting with $k = 0$, our CL-MP algorithm proceeds as follows. As is common in massive random access,  we assume that the noise power level $\sigma^2$ is known  and is given as input to the algorithm. 
  
{\bf Initialization}: Initialize $k=0$ and set 
\begin{itemize} 
\item The initial solution $\gam^{(0)}= \mathbf{0}_{N \times 1}$. 
\item The initial solution support  $\support^{(0)}= \mathsf{supp}(\gam^{(0)})=\emptyset$ 
\item The initial CM:  $\M^{(0)} =  \A \diag(\gam^{(0)}) \A^\hop + \sigma^{2} \mathbf{I} = \sigma^{2} \mathbf{I}$.  Then $(\M^{(0)})^{-1} = (1/\sigma^{2}) \bm{I} $ is the initial inverse CM. 
\end{itemize}    

{\bf Main Iteration} phase consists of the following steps: 

 {\it 1) Sweep:} Compute the errors 
\begin{align} \label{eq:epsilon_i_v0}
\epsilon_i  &=  \min_{\gamma \geq 0} \ell_i\big(\gamma\mid \M_{\setminus i}^{(k)} \big)  
\end{align}
for each $i \in \support^{\complement}=[\![\natoms]\!] \setminus \support^{(k)} $ 
using its unique optimal value 
\beq \label{eq:sweep_gamma_i}
\gamma_i  =  \max\Big( \frac{\a_i^\hop (\M^{(k)})^{-1} \S (\M^{(k)})^{-1} \a_i  - \a_i^\hop (\M^{(k)})^{-1} \a_i}{(\a_i^\hop (\M^{(k)})^{-1} \a_i)^2 } , 0\Big). 
\eeq

 {\it  2) Update support:}  Find a minimizer, $i_k$ of $\epsilon_i$: $\forall i \in  \support^{\complement}$, $\epsilon_{i_k} \leq \epsilon_i$, 
and update the support $\support^{(k+1)} = \support^{(k)} \cup \{ i_k\}$. 

{\it 3) Update the inverse CM}: Since the new covariance matrix after support update is 
\[
\M^{(k+1)}  = 
\M^{(k)} + \gamma_{i_k} \a_{i_k} \a_{i_k}^\hop, 
\]
its inverse covariance matrix $ ( \M^{(k+1)})^{-1}$ can be obtained using Sherman-Morrison\footnote{The Sherman-Morrison formula is only true if $\bm{\Sigma}$ is non singular and that the denominator $1+\bm{a}^\hop\bm{\Sigma}^{-1}\bm{a}\neq 0$. In case of the positive definite matrix, $\bm{\Sigma}$ all these conditions are fulfilled.} formula, which states that 
for $\M$ invertible, we have that 
\beq \label{eq:Minv_update}
(\M + \gamma \a \a^\hop)^{-1} =  \M^{-1}   -  \dfrac{ \gamma \M^{-1}   \a  \a^{\hop} \M^{-1}  }{ 1+ \gamma  \a^\hop \M^{-1} \a}
\eeq
 
 {\it  4) Stopping rule}: If stopping rule is not met, then increment $k$ by 1 and repeat steps 1)-3).  
 Several stopping rules can be used. For example, one can stop iterations after the fixed number (say $K_{\max}$) or when the added index has power level  $\gamma_{i_k}$ that falls below some threshold level.  In the former case, $K_{\max}$ serves as the maximum number of active devices expected to be active. 
  
It is still possible to reduce the computation times of the algorithm by not updating the inverse $\M^{-1}$ in step 3. Instead, $\B=\M^{-1} \A$ can be used. That is, one can replace the inverse CM update in step~3 
with the update of $\B$:  
\[
\B^{(k+1)}  =  [\M^{(k+1)}]^{-1}\A=  \B^{(k)}    -  \dfrac{ \gamma_{i_k} }{  1+ \gamma_{i_k}   \a_{i_k}^\hop \b_{i_k}^{(k)} } \b_{i_k}^{(k)}  \a_{i_k}^{\hop} \B^{(k)}, 
\]
which results  from multiplying  \eqref{eq:Minv_update} by $\A$ from the right.

Next we express $\ell_i(\gamma \mid \M_{\setminus i})$ in \eqref{eq:epsilon_i_v0_b} in more explicit  form. First, observe that 
\begin{align}
\log& | \M_{\setminus i} + \gamma \a_i \a_i^\hop |  \notag \\
&= \log |    \M_{\setminus i}^{1/2}(\boldsymbol{I} + \gamma  \M_{\setminus i}^{-1/2}\a_i \a_i^\hop \M_{\setminus i}^{-1/2} )\M_{\setminus i}^{1/2} | \notag  \\
&= \log |  \M_{\setminus i} |   + \log(1+ \gamma  \a_i^\hop \M_{\setminus i}^{-1}  \a_i) , \label{eq:epsilon_i_apu1}
\end{align}
where we applied Sylvester's determinant theorem\footnote{ It states that for $\A$, an $m \times n$ matrix, and $\B$, an $n \times m$ matrix, it holds that 
$| \bo I _m + \A \B | = | \bo I_n + \B \A |$.}. Using Sherman-Morrison formula \eqref{eq:Minv_update} we can write
\beq
 \tr( (\M_{\setminus i} + \gamma \a_i \a_i^\hop )^{-1}  \S)  =  \tr(\M_{\setminus i}^{-1} \hat \M) -  \gamma \frac{\a_i^\hop\M_{\setminus i}^{-1} \hat \M \M_{\setminus i}^{-1} \a_i}{1+\gamma \a_i \M_{\setminus i}^{-1} \a_i}. \label{eq:epsilon_i_apu2}
\eeq
Using \eqref{eq:epsilon_i_apu1} and \eqref{eq:epsilon_i_apu2} shows that  \eqref{eq:epsilon_i_v0_b} can be written as:
\beq \label{eq:epsilon_i_apu3}
\ell_i(\gamma \mid \M_{\setminus i})= c + \log(1+ \gamma  \a_i^\hop \M_{\setminus i}^{-1}  \a_i)  - \gamma \frac{\a_i^\hop\M_{\setminus i}^{-1} \hat \M \M_{\setminus i}^{-1} \a_i}{1+\gamma \a_i^\hop \M_{\setminus i}^{-1} \a_i},
\eeq
 where $c= \log |  \M_{\setminus i} |  +   \tr(\M_{\setminus i}^{-1} \hat \M)$ collects the irrelevant constant terms that are not dependent on $\gamma$. Thus, w.l.o.g., we  can set $c=0$. 

 Using \eqref{eq:epsilon_i_apu3} shows that the error term $\epsilon_i$ in \eqref{eq:epsilon_i_v0} has the following closed-form expression:
\beq \label{eq:epsilon_i}
\epsilon_i = \ell_i(\gamma_i  \mid \M_{\setminus i}^{(k)})= \log (1 + \gamma_i \a_i^\hop \b_i )  - \gamma_i  \frac{\b_i^\hop  \S  \b_i }{1+ \gamma_i \a_i^\hop \b_i},
\eeq 
where  $\gamma_i$ is the minimizer in \eqref{eq:sweep_gamma_i}, and where we have for simplicity of notation written $\B = \B^{(k)}$, i.e., we have dropped the superscript indexing. 
 If $\gamma_i$  in  \eqref{eq:sweep_gamma_i}  is non-zero,  meaning  $\gamma_i >0$, then it can be expressed compactly as:  
\[
\gamma_i  =  \frac{ \b_i^\hop \S  \b_i  -\a_i^\hop  \b_i }{ (\a_i^\hop \b_i)^2 } .
\]
This implies that 
\begin{align*}
 1+ \gamma_i \a_i^\hop  \b_i &= 1+  \frac{ \b_i^\hop \S  \b_i  -\a_i^\hop  \b_i }{ (\a_i^\hop \b_i)^2 } \a_i^\hop \b_i   \\
 &= 1 + \frac{\b_i^\hop \S  \b_i  - \a_i^\hop  \b_i}{\a_i^\hop \b_i} \\
 &= \frac{\b_i^\hop \S  \b_i}{\a_i^\hop \b_i}   \mbox{ when  $\gamma_i>0$}. 
 \end{align*} 
 Substituting this into the denominator of the last term in \eqref{eq:epsilon_i}, we obtain 
\beq \label{eq:epsilon_i_v2}
\epsilon_i = \log (1 + \gamma_i \a_i^\hop \b_i )  -  \gamma_i \a_i^\hop \b_i .
\eeq
In the case that  $\gamma_i = 0$, then the error $\epsilon_i$ in \eqref{eq:epsilon_i} becomes $\epsilon_i=0$. This explains line 3 in Algorithm~\ref{alg:CLMP}.

All the aforementioned considerations lead to the efficient implementations outlined in Algorithm~\ref{alg:CLMP}.  To promote reproducibility, an optimized MATLAB implementation of the method, along with scripts to replicate many of the simulation studies is available at~\url{https://github.com/esollila/CLMP}. 

\begin{remark} \label{rem:CLMPcomparison} It is also possible to derive an orthogonal matching pursuit  (OMP) for the problem at hand, following the developments in  \cite{ollila2024sparse}. The OMP algorithm includes an additional, ``update provisional solution" step, which occurs immediately after the ``update support" step. In this step, the signal power solutions are recomputed for the updated support. This involves solving:
\begin{align} \label{eq:optimize_gammas}
\hat \gam &= \underset{\gam \in \mathbb{R}^{N}_{\geq 0}}{\arg \min} \   \tr(  (\A   \Gam\A^\hop + \sigma^2 \bm{I})^{-1} \SCM )  +  \log |  \A \Gam \A^\hop + \sigma^2 \bm{I} |  \notag \\
		&\quad \mbox{subject to } \  \mathsf{supp}(\gam)=\support^{(k+1)}, \ \Gam = \diag(\gam)
 \end{align} 
The method for solving this  optimization problem is discussed in \cite{ollila2024sparse}. The OMP approach also requires significant modifications in the `update inverse covariance matrix' step. Specifically, after the signal powers are recalculated by solving \eqref{eq:optimize_gammas}, the inverse CM is updated as follows:
\[
(\M^{(k+1)}) ^{-1}  = (  \A \diag(\hat{\gam}) \A^\hop + \sigma^2 \bm{I} )^{-1} 
\]
which has a complexity of $\mathcal{O}(L^3)$, compared to the $\mathcal{O}(L^2)$ complexity in step 3 of the CL-MP algorithm, where the Sherman-Morrison formula is employed to compute the inverse.  Moreover, the 
CL-OMP approach does not allow for the use of the efficient update rule discussed earlier (updating $\B=\M^{-1} \A$).
In summary, compared with the CL-OMP algorithm, CL-MP benefits from a computationally efficient update rule,  where it updates $\B=\M^{-1}\A$ instead of the inverse CM, thereby speeding up the computations. Additionally, CL-MP avoids the need to solve the optimization problem \eqref{eq:optimize_gammas} required to update the signal power estimates. Given the computational constraints of massive random access, we believe that the MP approach is better suited for the task at hand. Our numerical simulations also confirmed that the CL-OMP approach yielded only marginal improvements in AD performance, while significantly increasing computation times compared to CL-MP.
\end{remark}

{\color{red} 
\begin{algorithm}[!h]
 \caption{\textbf{C}ovariance \textbf{L}earning \textbf{M}atching \textbf{P}ursuit (CL-MP) algorithm}\label{alg:CLMP}
\DontPrintSemicolon
\SetKwInOut{Input}{Input} 
\SetKwInOut{Output}{Output}
\SetKwInOut{Init}{Initialize}
\SetNlSkip{1em}
\SetInd{0.5em}{0.5em}
\Input{ $\S = \ndata^{-1} \bm{Y} \bm{Y}^\hop$, $\A$,  $\sigma^2$} 
\Init{ $\bm{B}=(1/\sigma^2)\bm{A}$, $\support=  \emptyset$}

 \For{$k =1,\ldots,\natoms$}{

$\gamma_i \gets \max\Big( \frac{\b_i^\hop  \S \b_i - \a_i^\hop \b_i }{(\a_i^\hop \b_i)^2 }, 0  \Big)$, $ \forall i \in \support^{\complement}$

\BlankLine
$\epsilon_i   \gets   \log (1 + \gamma_i \a_i^\hop \b_i )  -  \gamma_i \a_i^\hop \b_i $, $\forall i \in \support^{\complement}$

\BlankLine
 $\support   \gets \support \cup \{ i_k \}$ with $i_k  \gets   \arg \min_{i \in \support^{\complement}} \epsilon_i $

\BlankLine
\If{stopping rule is met (see text)}{\nonl \textsf{break}}

\BlankLine
 $\B  \gets \B   -  \dfrac{ \gamma_{i_k} }{  1+ \gamma_{i_k}   \a_{i_k}^\hop \b_{i_k} } \b_{i_k}  \a_{i_k}^{\hop} \B $

   }
 \Output{set $\support$,  containing $I$ indices, where $I$ is the
number of iterations completed.} 
\end{algorithm} }









\subsection{Computational complexity comparisons}

In Table~\ref{complexAnalysis}, we present the computational complexity analysis of CL-MP and some benchmark algorithms. In the complexity calculations, it is assumed that the 
number of antennas ($M$) is smaller than the number of users ($N$) and the pilot length ($L$) is smaller than $N$, which are 
typical scenarios in mMTC. In SOMP, the index $k$ denotes the iteration index (i.e., the number of atoms selected). 
The dominant complexity of CL-MP Algorithm~\ref{alg:CLMP} is the matrix-vector
multiplications in lines 2, 3, and 7. The complexity of lines 2 and 3 is $\mathcal{O}(L^2)$. Since these are computed for (essentially) all $N$ devices, the total complexity of these steps is $\mathcal{O}(NL^2)$. In line 4, one searches the smallest value in a set of $N$ scalars with linear complexity $\mathcal O(N)$. 
The complexity of line 7 is $\mathcal O(NL)$. 
Hence,  the full complexity of Algorithm 1 per iteration is $\mathcal{O}(NL^2)$.  
Thus the complexity of CL-MP is linear in $N$ and quadratic in $L$. This is a desired feature for low-latency mMTC scenarios where the number of MTDs $N$ is large while the pilot length $L$ is often small. Note that SBL and VAMP grow quadratic in $N$.

 It is also worth noting that the computational complexity of CL-MP per iteration is the same as that of CWO.  However, CL-MP follows a greedy approach, wheras CWO is an iterative method that achieves its best performance only after convergence. Unlike iterative approaches, greedy methods are not affected by  convergence issues, which can arise when the nominal assumptions such as data  Gaussianity do not hold or when  outliers are present. 
 
It is also important to emphasize that the overall complexity of all algorithms is determined by the complexity per iteration multiplied by the number of iterations $I$ required for convergence. For instance, the total complexity of SBL is $\mathcal O(I N^2 L)$. It is well known that the Expectation Maximization (EM) algorithm utilized by SBL suffers from slow convergence, and thus $I$ can often be very large. In contrast, greedy pursuit approaches do not suffer from this issue, as they aim to find the minimum of the LLF greedily rather than relying on an iterative optimization algorithm. 

{\begin{table}[t!]
\centering
\caption{Computational complexity of  CS-MUD algorithms.}
\small
\begin{tabularx}{0.35\textwidth}{c||c} 
 \hline
  \textbf{Algorithm} &  \textbf{Complexity (per iteration)} \\ [0.7ex] 
 \hline\hline
CL-MP & $\mathcal{O}(NL^2)$\\
CWO \cite{fengler2021non}  &  $\mathcal{O}(NL^2)$ \\
VAMP \cite{liu2018massive} & $\mathcal{O}(N^2M)$\\  
SBL \cite{wipf2004sparse} & $\mathcal{O}(N^2L)$ \\
SOMP  \cite{tropp2006algorithms}  & $O(L  \cdot \max(N M, k^2) + k^3)$\\ 
 \hline
\end{tabularx}
\label{complexAnalysis}
\end{table}}

\section{Empirical comparisons}
\label{Results}
In this section, we evaluate the performance of the proposed CL-MP algorithm under a variety of settings.  
As performance metrics, we consider probability of misdetection (PMD)  and exact recovery rate (ERR) . {\it Misdetection} occurs when an active device
is incorrectly classified as inactive. The (empirical) PMD is computed as the Monte Carlo (MC) average ratio of the ratio of misdetected devices to the total number of active MTDs, defined as 
\begin{equation}
      \text{PMD} = \frac{1}{Q} \sum_{q=1}^Q \frac{| \mathcal K \setminus \mathcal K^{(q)}| }{| \mathcal K |},
\end{equation}
where $Q$ is the number of MC trials, and $\mathcal K^{(q)}$ is the estimate of the support $\mathcal K$ of the $q^{\text{th}}$ MC trial. On the other hand, {\it exact recovery} of devices occurs when all active devices are correctly identified, and it is defined as
\begin{equation}
   \text{ERR}  =  \frac{1}{Q} \sum_{q=1}^Q \mathbbm{1}( \mathcal{K}^{(q)} = \mathcal K).
\end{equation}
with  $\mathcal{K}$ representing the set of active device indices with $|\mathcal{K}| = K$, and for each q$^{\text{th}}$ MC trial, these indices are randomly selected from the set $[\![ N ]\!]$ without replacement.
We note that the indices of active devices $\mathcal K$  (with  $|\mathcal K |= K$) for each $q^{\text{th}}$ MC trial are also randomly selected from set $[\! [  N  ] \!]$ without replacement. 

\subsection{The effect of $M$, $L$ and $K$}
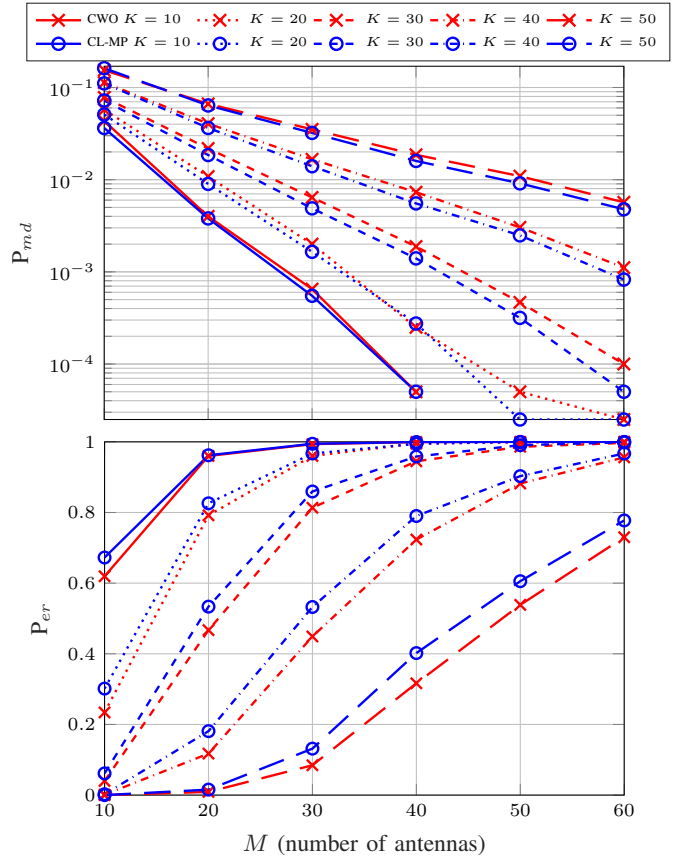
\begin{figure}[!t]
\centerline{\hspace{3pt}\begin{tikzpicture}

\begin{axis}[%
width=0.78\columnwidth,
height=4.7cm,
scale only axis,
scale only axis,
xmin=10,
xmax=60,
xlabel style={font=\small\color{white!15!black}},
tick label style={font=\scriptsize} ,
ylabel={PMD},
ymode=log,
ymin=0.000025,
ymax=0.17,
yminorticks=true,
xmajorgrids,
ymajorgrids,
yminorgrids,
xticklabel=\empty,
ylabel style={font=\small\color{white!15!black}},
axis background/.style={fill=white},
title style={font=\bfseries},
legend style={at={(-0.15,1.18)}, anchor=north west, legend cell align=left, align=left, draw=white!15!black,font=\tiny,legend columns=5}
]
\addplot [color=red, line width=0.9pt, mark size=3.3, mark=x, mark options={solid, red}]
  table[row sep=crcr]{%
10	0.0433499999999998\\
20	0.00399999999999999\\
30	0.00065\\
40	5e-05\\
50	0\\
60	0\\
};
\addlegendentry{CWO $K=10$}

\addplot [color=red, dotted, line width=0.9pt, mark size=3.3, mark=x, mark options={solid, red}]
  table[row sep=crcr]{%
10	0.0574249999999988\\
20	0.0108250000000001\\
30	0.002\\
40	0.00025\\
50	5e-05\\
60	2.5e-05\\
};
\addlegendentry{$K=20$}

\addplot [color=red, dashed, line width=0.9pt, mark size=3.3, mark=x, mark options={solid, red}]
  table[row sep=crcr]{%
10	0.0780333333333319\\
20	0.0218000000000001\\
30	0.00641666666666665\\
40	0.00188333333333333\\
50	0.000466666666666667\\
60	0.0001\\
};
\addlegendentry{$K=30$}

\addplot [color=red, dashdotted, line width=0.9pt, mark size=3.3, mark=x, mark options={solid, red}]
  table[row sep=crcr]{%
10	0.113087499999999\\
20	0.0405499999999998\\
30	0.0166249999999998\\
40	0.00731250000000007\\
50	0.00303750000000001\\
60	0.0011125\\
};
\addlegendentry{$K=40$}

\addplot [color=red,  line width=0.9pt, mark size=3.3, mark=x, mark options={solid, red},dash pattern=on 10pt off 5pt on 10pt off 5pt]
  table[row sep=crcr]{%
10	0.15413\\
20	0.0664500000000005\\
30	0.0351200000000003\\
40	0.0186199999999999\\
50	0.0109199999999998\\
60	0.00567999999999992\\
};
\addlegendentry{$K=50$}

\addplot [color=blue, line width=0.9pt, mark size=2.3, mark=o, mark options={solid, blue}]
  table[row sep=crcr]{%
10	0.0363500000000001\\
20	0.00379999999999999\\
30	0.00055\\
40	5e-05\\
50	0\\
60	0\\
};
\addlegendentry{CL-MP $K=10$}

\addplot [color=blue, dotted, line width=0.9pt, mark size=2.3, mark=o, mark options={solid, blue}]
  table[row sep=crcr]{%
10	0.050949999999999\\
20	0.00900000000000006\\
30	0.00165\\
40	0.000275\\
50	2.5e-05\\
60	2.5e-05\\
};
\addlegendentry{$K=20$}

\addplot [color=blue, dashed, line width=0.9pt, mark size=2.3, mark=o, mark options={solid, blue}]
  table[row sep=crcr]{%
10	0.0722833333333322\\
20	0.0185166666666669\\
30	0.00488333333333332\\
40	0.0014\\
50	0.000316666666666667\\
60	5e-05\\
};
\addlegendentry{$K=30$}

\addplot [color=blue, dashdotted, line width=0.9pt, mark size=2.3, mark=o, mark options={solid, blue}]
  table[row sep=crcr]{%
10	0.1107625\\
20	0.0363999999999998\\
30	0.0139874999999999\\
40	0.00552500000000004\\
50	0.0024875\\
60	0.000824999999999999\\
};
\addlegendentry{$K=40$}

\addplot [color=blue, line width=0.9pt, mark size=2.3, mark=o, mark options={solid, blue},dash pattern=on 10pt off 5pt on 10pt off 5pt]
  table[row sep=crcr]{%
10	0.1619\\
20	0.0638700000000004\\
30	0.0321600000000004\\
40	0.0159699999999997\\
50	0.00911999999999987\\
60	0.00476999999999995\\
};
\addlegendentry{$K=50$}

\end{axis}

\end{tikzpicture}%
}
\vspace{-4pt}
\centerline{\begin{tikzpicture}

\begin{axis}[%
width=0.78\columnwidth,
height=4.7cm,
scale only axis,
xmin=10,
xmax=60,
xmajorgrids,
ymajorgrids,
tick label style={font=\scriptsize} ,
xlabel style={font=\small\color{white!15!black}},
xlabel={$M$ (number of antennas)},
ymin=0,
ymax=1,
ylabel style={font=\small\color{white!15!black}},
ylabel = {ERR}, 
axis background/.style={fill=white},
legend style={legend cell align=left, align=left, draw=white!15!black}
]
\addplot [color=red, line width=0.9pt, mark size=3.0pt, mark=x, mark options={solid, red}]
  table[row sep=crcr]{%
10	0.6195\\
20	0.96\\
30	0.9935\\
40	0.9995\\
50	1\\
60	1\\
};

\addplot [color=red, dotted, line width=0.9pt, mark size=3.0pt, mark=x, mark options={solid, red}]
  table[row sep=crcr]{%
10	0.234\\
20	0.792\\
30	0.96\\
40	0.995\\
50	0.999\\
60	0.9995\\
};

\addplot [color=red, dashed, line width=0.9pt, mark size=3.0pt, mark=x, mark options={solid, red}]
  table[row sep=crcr]{%
10	0.04\\
20	0.467\\
30	0.8135\\
40	0.945\\
50	0.986\\
60	0.997\\
};

\addplot [color=red, dashdotted, line width=0.9pt, mark size=3.0pt, mark=x, mark options={solid, red}]
  table[row sep=crcr]{%
10	0.001\\
20	0.1175\\
30	0.449\\
40	0.7235\\
50	0.882\\
60	0.956\\
};

\addplot [color=red, line width=0.9pt, mark size=3.0pt, mark=x, mark options={solid, red},dash pattern=on 10pt off 5pt on 10pt off 5pt]
  table[row sep=crcr]{%
10	0\\
20	0.009\\
30	0.0845\\
40	0.3165\\
50	0.5385\\
60	0.73\\
};

\addplot [color=blue, line width=0.9pt, mark size=2.2pt, mark=o, mark options={solid, blue}]
  table[row sep=crcr]{%
10	0.6725\\
20	0.962\\
30	0.9945\\
40	0.9995\\
50	1\\
60	1\\
};

\addplot [color=blue, dotted, line width=0.9pt, mark size=2.2pt, mark=o, mark options={solid, blue}]
  table[row sep=crcr]{%
10	0.3015\\
20	0.826\\
30	0.967\\
40	0.9945\\
50	0.9995\\
60	0.9995\\
};

\addplot [color=blue, dashed, line width=0.9pt, mark size=2.2pt, mark=o, mark options={solid, blue}]
  table[row sep=crcr]{%
10	0.0615\\
20	0.5335\\
30	0.8595\\
40	0.9585\\
50	0.9905\\
60	0.9985\\
};

\addplot [color=blue, dashdotted, line width=0.9pt, mark size=2.2pt, mark=o, mark options={solid, blue}]
  table[row sep=crcr]{%
10	0.0025\\
20	0.181\\
30	0.5325\\
40	0.79\\
50	0.903\\
60	0.967\\
};

\addplot [color=blue, line width=0.9pt, mark size=2.2pt, mark=o, mark options={solid, blue},dash pattern=on 10pt off 5pt on 10pt off 5pt]
  table[row sep=crcr]{%
10	0\\
20	0.0155\\
30	0.1315\\
40	0.402\\
50	0.6055\\
60	0.7775\\
};

\end{axis}
\end{tikzpicture}%
}
\caption{Probability of misdetection  (top) and exact recovery rate (bottom) vs $M$ for different levels of active devices ($K$); $L=64$, $N=1000$, and  LSFC-s $\beta_n\sim Unif(-15, 0)$ on dB scale. } \label{fig1}
\end{figure}
We first investigate the effects of $L$, $M$, and $K$ on the 
AUD performance. 
We let the LSFCs $\beta_n$, $\forall n$ to follow   {\it uniform distribution} in dB scale, i.e.,  $10\log_{10}(\beta_n)$ is uniformly distributed in range $[10\log_{10}(\beta_\text{min}),10\log_{10}(\beta_\text{max})]$. 
We compare the proposed CL-MP in Algorithm~\ref{alg:CLMP} against the CWO algorithm  \cite[Algorithm~1]{fengler2021non}.  We set the maximum number of iterations of the CWO algorithm to 15.  
To ensure a fair comparison, we assume that the number of active devices, $K$, is known to both algorithms. Hence the CL-MP algorithm is terminated after $K$ iterations and directly returns the estimated support $\hat{\support}$ of size $|\hat{\support}|= K$. The CWO algorithm requires thresholding, where the active users are identified by picking the positions of the $K$ largest entries of the final estimate $\hat{\bm{\gamma}}$.  The number of MC runs is $2000$.

\begin{figure}[!t]
\centerline{\definecolor{mycolor1}{rgb}{0.00000,0.44700,0.74100}%
\definecolor{mycolor2}{rgb}{0.85000,0.32500,0.09800}%
\begin{tikzpicture}

\begin{axis}[%
width=0.82\columnwidth,
height=2.5cm,
scale only axis,
bar shift auto,
label style={font=\small\color{white!15!black}},
ylabel = {Runtime (s)},
xlabel = {$K$ (number of active devices)}, 
xmin=5.14285714285714,
xmax=54.8571428571429,
xtick={10, 20, 30, 40, 50},
tick label style={font=\scriptsize} ,
yticklabel=\pgfkeys{/pgf/number format/.cd,fixed,precision=2,zerofill}\pgfmathprintnumber{\tick},
ymin=0,
ymax=0.12,
axis background/.style={fill=white},
legend style={at={(0.03,1.0)}, anchor=north west, legend cell align=left, legend columns=2, align=left, draw=white!15!black,font=\scriptsize,opacity=0.99}
]
\addplot[ybar, bar width=2.286, fill=mycolor1, draw=black, area legend] table[row sep=crcr] {%
10	0.0203390221255556\\
20	0.0405168592122222\\
30	0.0607953093166667\\
40	0.0811169719588889\\
50	0.101295243313333\\
};
\addplot[forget plot, color=white!15!black] table[row sep=crcr] {%
5.14285714285714	0\\
54.8571428571429	0\\
};
\addlegendentry{CL-MP}

\addplot[ybar, bar width=2.286, fill=mycolor2, draw=black, area legend] table[row sep=crcr] {%
10	0.0816522411122222\\
20	0.0888650814311111\\
30	0.0964707973711111\\
40	0.103698483561111\\
50	0.112063790152222\\
};
\addplot[forget plot, color=white!15!black] table[row sep=crcr] {%
5.14285714285714	0\\
54.8571428571429	0\\
};
\addlegendentry{CWO}

\end{axis}
\end{tikzpicture}%
}
\centerline{\definecolor{mycolor1}{rgb}{1.00000,0.00000,1.00000}%
\begin{tikzpicture}
\begin{groupplot}[group style={group size=4 by 1,horizontal sep=0.29cm,vertical sep=0.27cm}]

\nextgroupplot[
width=0.42\columnwidth,
height=3.0cm,
scale only axis,
xmin=30,
xmax=80,
xlabel style={font=\small\color{white!15!black}},
tick label style={font=\scriptsize} , 
xlabel={$L$ (pilot length)},
ymin=0,
ymax=0.18,
ylabel style={font=\small\color{white!15!black}},
yticklabel=\pgfkeys{/pgf/number format/.cd,fixed,precision=2,zerofill}\pgfmathprintnumber{\tick},
ylabel={Runtime (s)},
axis background/.style={fill=white},
title style={font=\bfseries},
title={CWO},
xmajorgrids,
ymajorgrids,
legend style={legend cell align=left, align=left, draw=white!15!black}
]
\addplot [color=red, line width=0.8pt, mark size=2.5pt, mark=x, mark options={solid, red}]
  table[row sep=crcr]{%
32	0.026105593902\\
48	0.048338807708\\
64	0.077325227522\\
80	0.120641497832\\
};

\addplot [color=blue, line width=0.8pt, mark size=2.5pt, mark=x, mark options={solid, blue}]
  table[row sep=crcr]{%
32	0.029936991946\\
48	0.054432399148\\
64	0.084992762888\\
80	0.131380804002\\
};

\addplot [color=black, line width=0.8pt, mark size=2.5pt, mark=x, mark options={solid, black}]
  table[row sep=crcr]{%
32	0.032055204982\\
48	0.061886829666\\
64	0.09305093756\\
80	0.141011982286\\
};

\addplot [color=mycolor1, line width=0.8pt, mark size=2.5pt, mark=x, mark options={solid, mycolor1}]
  table[row sep=crcr]{%
32	0.033901457846\\
48	0.068783722766\\
64	0.103392064622\\
80	0.151450588428\\
};

\addplot [color=green, line width=0.8pt, mark size=2.5pt, mark=x, mark options={solid, green}]
  table[row sep=crcr]{%
32	0.035633004094\\
48	0.072749981976\\
64	0.11393333642\\
80	0.163612604252\\
};

\nextgroupplot[
width=0.42\columnwidth,
height=3.0cm,
scale only axis,
xmin=30,
xmax=80,
yticklabel=\empty,
xlabel style={font=\small\color{white!15!black}},
tick label style={font=\scriptsize} , 
xlabel={$L$ (pilot length)},
ymin=0,
ymax=0.18,
axis background/.style={fill=white},
title style={font=\bfseries},
title={CL-MP},
xmajorgrids,
ymajorgrids,
legend style={at={(0.0,1)}, anchor=north west, legend cell align=left, align=left, draw=white!15!black,font=\scriptsize,legend columns=1,opacity=0.8}
]
\addplot [color=red, line width=0.8pt, mark size=2.5pt, mark=x, mark options={solid, red}]
  table[row sep=crcr]{%
32	0.00744016203\\
48	0.013186832878\\
64	0.02032241715\\
80	0.029304055888\\
};
\addlegendentry{$K=10$}

\addplot [color=blue, line width=0.8pt, mark size=2.5pt, mark=x, mark options={solid, blue}]
  table[row sep=crcr]{%
32	0.014602381756\\
48	0.02606291533\\
64	0.040487816754\\
80	0.058459851674\\
};
\addlegendentry{$K=20$}

\addplot [color=black, line width=0.8pt, mark size=2.5pt, mark=x, mark options={solid, black}]
  table[row sep=crcr]{%
32	0.021932439618\\
48	0.038778601254\\
64	0.060910113932\\
80	0.087780328538\\
};
\addlegendentry{$K=30$}

\addplot [color=mycolor1, line width=0.8pt, mark size=2.5pt, mark=x, mark options={solid, mycolor1}]
  table[row sep=crcr]{%
32	0.0293183895\\
48	0.052133402398\\
64	0.081185297842\\
80	0.116989981738\\
};
\addlegendentry{$K=40$}

\addplot [color=green, line width=0.8pt, mark size=2.5pt, mark=x, mark options={solid, green}]
  table[row sep=crcr]{%
32	0.036536069338\\
48	0.06483882875\\
64	0.101609284444\\
80	0.146141901314\\
};
\addlegendentry{$K=50$}

\end{groupplot}
\end{tikzpicture}%
}
\caption{Top: Average running  time in seconds for different $K$ when $\dim=64$ and $\natoms=1000$. Bottom: Average running time 
 for different $L$ and $K$  when $\ndata=40$ for CWO (left panel) and CL-MP (right panel) methods.}  \label{fig:CPU}
\end{figure}
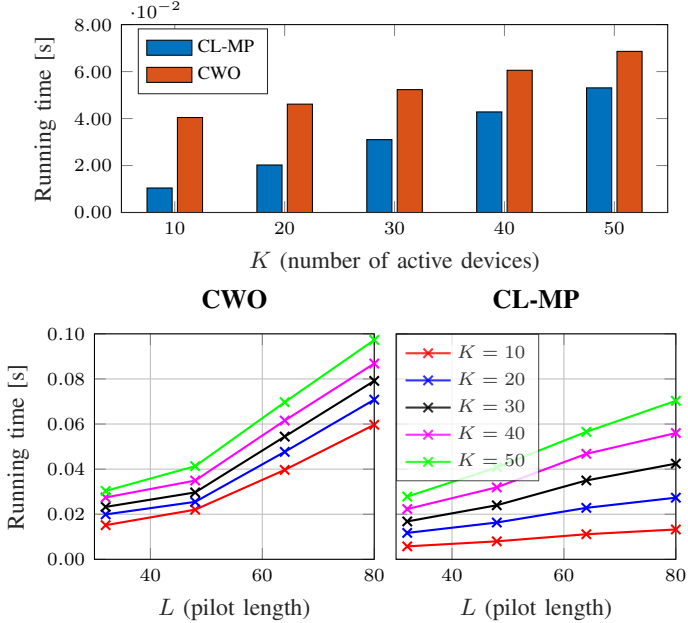

 \autoref{fig1} displays the probability of misdetection and exact recovery rate as a function of the number of antennas ($\ndata$)  and different numbers of active devices ($K$).  Bernoulli pilot sequence of length $\dim=64$ are used and the number of devices is $N=1000$. The pilots  are normalized to unit energy per symbol, i.e.,  
$\|\a_n\|_2^2=\dim$. LSFC coefficients follow a uniform distribution with a range $[-15, 0]$ dB. Both methods show similar performance, but the CL-MP has a slight advantage in all levels of $K$ and $M$. In addition to this, the running times (on an MacBook Pro M3 laptop) are shown in the top panel of \autoref{fig:CPU} demonstrate that it achieves high-level performance in significantly lower running time than the CWO algorithm. To conduct such comparison properly, we closed all background processes in
our computer (internet, e-mail, music,  unrelated software), and restricted MATLAB to use a single processor all time. It is worth noting that when allowing the use of all 8 cores in the laptop, the CL-MP method was significantly (namely, one order of magnitude) faster than CWO. 
This can be explained by the fact that CL-MP can compute essentially all lines in the algorithm using just matrix-vector multiplications.
We also note that the CWO implementation used random permutation strategy (which randomly permutes all coordinates
and then updates the coordinate one by one according to the order in the permutation) which has been reported to be more efficient strategy for computing the CWO estimate \cite{liu2024grant}\footnote{We used the efficient MATLAB implementation of the original authors available at \url{https://github.com/AlexFengler/mmv-activity-and-random-access}}. As observed, the differences between the methods become more pronounced when the number of active devices is small. Since both the CWO and CL-MP algorithms do not scale with $\ndata$, the reported CPU  times represent averages over all instances of $\ndata$. This further highlights the computational efficiency of the proposed method. 
 
  \autoref{fig:LvsAD} displays the probability of misdetection and exact recovery rate as a function of the pilot sequence length ($\dim$)  and the number of active devices ($K$) when the number of antennas is fixed ($\ndata=40$).  Again, we can note that the proposed  CL-MP performed better for all levels of $K$ and $L$. The running times (on a MacBook Pro M3 laptop) for both methods are displayed at the bottom panel of \autoref{fig:CPU}. As noted, CL-MP scales better with $L$ than CWO, and the advantage is, as expected, more significant when $K$ is small.  

\begin{figure}[!t]
\centerline{\hspace{2pt}\begin{tikzpicture}

\begin{axis}[%
width=0.78\columnwidth,
height=4.8cm,
scale only axis,
xmin=32,
xmax=80,
xlabel style={font=\color{white!15!black}},
tick label style={font=\scriptsize} ,
label style={font=\small\color{white!15!black}},
ymode=log,
ymin=3e-05,
ymax=0.31,
yminorticks=true,
ylabel style={font=\color{white!15!black}},
ylabel = {PMD}, 
axis background/.style={fill=white},
xtick = {32, 48,    64,    80}, 
xmajorgrids,
ymajorgrids,
yminorgrids,
xticklabel=\empty,
legend style={at={(-0.15,1.18)}, anchor=north west, legend cell align=left, align=left, draw=white!15!black,font=\tiny,legend columns=5}
]
\addplot [color=red, line width=0.9pt, mark size=3.3pt, mark=x, mark options={solid, red}]
  table[row sep=crcr]{%
32	0.0257500000000002\\
48	0.00125\\
64	0.0001\\
80	0\\
};
\addlegendentry{CWO $K=10$}

\addplot [color=red, dotted, line width=0.9pt, mark size=3.3pt, mark=x, mark options={solid, red}]
  table[row sep=crcr]{%
32	0.0670749999999988\\
48	0.00627500000000002\\
64	0.000375\\
80	0\\
};
\addlegendentry{$K=20$}

\addplot [color=red, dashed, line width=0.9pt, mark size=3.3pt, mark=x, mark options={solid, red}]
  table[row sep=crcr]{%
32	0.13023333333333\\
48	0.0223166666666667\\
64	0.00163333333333333\\
80	6.66666666666667e-05\\
};
\addlegendentry{$K=30$}

\addplot [color=red, dashdotted, line width=0.9pt, mark size=3.3pt, mark=x, mark options={solid, red}]
  table[row sep=crcr]{%
32	0.206925\\
48	0.0497250000000001\\
64	0.00695000000000006\\
80	0.0004375\\
};
\addlegendentry{$K=40$}

\addplot [color=red, line width=0.9pt, mark size=3.3pt, mark=x, mark options={solid, red},dash pattern=on 10pt off 5pt on 10pt off 5pt]
  table[row sep=crcr]{%
32	0.272749999999999\\
48	0.0926100000000001\\
64	0.01921\\
80	0.00236999999999999\\
};
\addlegendentry{$K=50$}

\addplot [color=blue, line width=0.9pt, mark size=2.3pt, mark=o, mark options={solid, blue}]
  table[row sep=crcr]{%
32	0.0208000000000001\\
48	0.00085\\
64	5e-05\\
80	0\\
};
\addlegendentry{CL-MP $K=10$}

\addplot [color=blue, dotted, line width=0.9pt, mark size=2.3pt, mark=o, mark options={solid, blue}]
  table[row sep=crcr]{%
32	0.0600749999999988\\
48	0.0048\\
64	0.00035\\
80	0\\
};
\addlegendentry{$K=20$}

\addplot [color=blue, dashed, line width=0.9pt, mark size=2.3pt, mark=o, mark options={solid, blue}]
  table[row sep=crcr]{%
32	0.128716666666664\\
48	0.0187166666666668\\
64	0.00135\\
80	8.33333333333333e-05\\
};
\addlegendentry{$K=30$}

\addplot [color=blue, dashdotted, line width=0.9pt, mark size=2.3pt, mark=o, mark options={solid, blue}]
  table[row sep=crcr]{%
32	0.220724999999999\\
48	0.046625\\
64	0.00581250000000004\\
80	0.0004125\\
};
\addlegendentry{$K=40$}

\addplot [color=blue, line width=0.9pt, mark size=2.3pt, mark=o, mark options={solid, blue},dash pattern=on 10pt off 5pt on 10pt off 5pt]
  table[row sep=crcr]{%
32	0.304229999999998\\
48	0.0929500000000003\\
64	0.0164699999999998\\
80	0.00185\\
};
\addlegendentry{$K=50$}

\end{axis}
\end{tikzpicture}%
}
\vspace{-4pt}
\centerline{\begin{tikzpicture}

\begin{axis}[%
width=0.78\columnwidth,
height=4.8cm,
tick label style={font=\scriptsize} ,
label style={font=\small\color{white!15!black}},
scale only axis,
xmin=30,
xmax=80,
xlabel style={font=\color{white!15!black}},
xlabel={$L$ (pilot length)},
ymin=0,
ymax=1,
ylabel style={font=\color{white!15!black}},
ylabel = {ERR}, 
axis background/.style={fill=white},
xmajorgrids,
ymajorgrids,
legend style={legend cell align=left, align=left, draw=white!15!black}
]
\addplot [color=red, line width=0.9pt, mark size=3.3pt, mark=x, mark options={solid, red}]
  table[row sep=crcr]{%
32	0.757\\
48	0.9875\\
64	0.999\\
80	1\\
};

\addplot [color=red, dotted, line width=0.9pt, mark size=3.3pt, mark=x, mark options={solid, red}]
  table[row sep=crcr]{%
32	0.1835\\
48	0.876\\
64	0.9925\\
80	1\\
};

\addplot [color=red, dashed, line width=0.9pt, mark size=3.3pt, mark=x, mark options={solid, red}]
  table[row sep=crcr]{%
32	0.0015\\
48	0.4525\\
64	0.9515\\
80	0.998\\
};

\addplot [color=red, dashdotted, line width=0.9pt, mark size=3.3pt, mark=x, mark options={solid, red}]
  table[row sep=crcr]{%
32	0\\
48	0.0615\\
64	0.7415\\
80	0.9825\\
};

\addplot [color=red, line width=0.9pt, mark size=3.3pt, mark=x, mark options={solid, red},dash pattern=on 10pt off 5pt on 10pt off 5pt]
  table[row sep=crcr]{%
32	0\\
48	0.001\\
64	0.302\\
80	0.886\\
};

\addplot [color=blue, line width=0.9pt, mark size=2.3pt, mark=o, mark options={solid, blue}]
  table[row sep=crcr]{%
32	0.804\\
48	0.9915\\
64	0.9995\\
80	1\\
};

\addplot [color=blue, dotted, line width=0.9pt, mark size=2.3pt, mark=o, mark options={solid, blue}]
  table[row sep=crcr]{%
32	0.239\\
48	0.9055\\
64	0.993\\
80	1\\
};

\addplot [color=blue, dashed, line width=0.9pt, mark size=2.3pt, mark=o, mark options={solid, blue}]
  table[row sep=crcr]{%
32	0.003\\
48	0.5355\\
64	0.96\\
80	0.9975\\
};

\addplot [color=blue, dashdotted, line width=0.9pt, mark size=2.3pt, mark=o, mark options={solid, blue}]
  table[row sep=crcr]{%
32	0\\
48	0.0835\\
64	0.782\\
80	0.9835\\
};

\addplot [color=blue, line width=0.9pt, mark size=2.3pt, mark=o, mark options={solid, blue},dash pattern=on 10pt off 5pt on 10pt off 5pt]
  table[row sep=crcr]{%
32	0\\
48	0.002\\
64	0.3845\\
80	0.9095\\
};

\end{axis}
\end{tikzpicture}%
}
\caption{Probability of misdetection  (top) and exact recovery rate (bottom) vs $L$ for different levels of active devices ($K$); $N=1000$, $\ndata=40$ and LSFC-s $\beta_n \sim Unif(-15, 0)$ on dB scale.} \label{fig:LvsAD}
\end{figure}


 \subsection{MIMO uplink network simulation} 
 In this subsection, we present the results of the performance of our proposed CL-MP in the mMTC network. For these simulations, we consider a single-cell massive MIMO uplink network, where a BS serves $1000$ MTDs randomly placed within the cell between a radius of $25$ m and $250$ m. 
 We adopt a log-distance path loss model where LSFC is $\beta_{n}=-130-37.6\log_{10} d_{n}$, where $d_{n}$ is the distance between the $n^{\text{th}}$ MTD and the BS. For resource fairness, we assume uplink power control following channel inversion, such that each MTD transmits with power\cite{senel2018grant} 
 \begin{equation}
   \rho_{n} = \frac{p^{\mathrm{max}}\beta_{\mathrm{min}}}{\beta_n},
\end{equation} 
where $\beta_{\mathrm{min}}$ is the large-scale effect at the edge of the cell and $p^{\mathrm{max}}$ is the maximum allowable power for each MTD. Consequently, the received signal-to-noise ratio (SNR) for each MTD is defined by 
\begin{equation}    \text{SNR}_{n}=\frac{p^{\mathrm{max}}\beta_{\mathrm{min}}}{\sigma^2}.
    \label{snrValues}
\end{equation}
Unless otherwise stated, the simulation parameters used in this paper are summarized in Table~\ref{simulationPar}. \par For the performance benchmark, we compare CL-MP with Vector AMP (VAMP) \cite{liu2018massive}, SBL \cite{wipf2004sparse},  
and SOMP \cite{tropp2006algorithms}. We acknowledge that all these benchmark solutions first estimate $\X$, then map it to $\hat{\boldsymbol{\alpha}}$. To ensure fairness in our comparison, we input the number of active MTDs $K$ into each algorithm. This ensures that $K$ indices corresponding to maximum row norms are selected from matrix $\bm{X}$ after the convergence of each algorithm. Therefore, the set of indices of the selected $K$ rows will serve as an estimate $\hat{\mathcal{K}}$ of $\mathcal{K}$ at the end of each algorithm. Each algorithm will have two stopping criteria, i.e. if the maximum number of iterations $T_{max}$ has been reached or the relative change between two consecutive estimates $\Delta$ becomes insignificant.  For random sensing matrices, we use the Gaussian pilot sequences, which we denote by G-MAT, and the Bernoulli pilot sequences, which we denote by B-MAT. For structured pilot sequences, we generate pilot sequences following the restrictively combined fixed orthogonal block matrix from \cite{marata2023joint}, which we denoted by RC-FOBM as well as restively combined random orthogonal block matrix, which we denote by RC-ROBM.

\begin{table}
    \centering
  \caption{Simulation parameters}
 \begin{tabular}{  m{14em}  m{10em} } 
\hline
\textbf{Parameter} & \textbf{Value}\\ 
\hline
Cell radius & $250$ m \\ 
Number of MTDs $(N)$ & $1000$ \\
Bandwidth  & $20$ MHz  \\ 
Noise power ($\sigma^2$) & $2\!\times\!10^{-13}$ W \\ 
Coherence interval $(T)$ & $300$ \\ 
Length of the pilot sequences $(L)$ & $64$ \\ 
Number of BS antennas $(M)$ & $32$ \\ 
Activation probability $(\epsilon)$ & $0.01$ \\ 
$\text{Maximum MTD power}(p^{\text{max}})$ & $0.1$ W \\ 
$\text{Average SNR}$ & $10$ dB \\ 
Error tolerance $(\Delta)$ & $10^{-6}$ \\
Monte Carlo $(\text{MC})$ & $2000$ \\
Maximum Iterations $(T_{max})$ & $150$ \\
\hline
\end{tabular}
\label{simulationPar}
\end{table}

Fig.~\ref{pmdVsSNRcompare} presents the results in terms of the PMD as a function of the SNR, for different algorithms. The general trend shows that increasing SNR improves the PMD of all the algorithms. This is because as the SNR increases, the desired signal component, which in our case is represented by $\bm{A}\bm{X}$, has higher power relative to the noise. As a result, it becomes easier to recover and identify the active users. The results show that both SBL and CL-MP have comparable performance as the SNR increases and their performance is superior to that of the SOMP and VAMP. The comparable performance between SBL and CL-MP is due to the equivalence of their objective functions as mentioned in Section~\ref{contributions}. However, the superior performance of CL-MP validates the proposed solutions. Since CL-MP has a similar performance as SBL, we next evaluate its performance for different sensing matrices. 
\begin{figure}
    \centering    
%
%
\definecolor{mycolor1}{rgb}{0.54510,0.00000,0.54510}%
\definecolor{mycolor2}{rgb}{0.00000,0.54510,0.54510}%
\definecolor{mycolor3}{rgb}{0.00000,0.00000,0.50196}%
\definecolor{mycolor4}{rgb}{0.86275,0.07843,0.23529}%
\definecolor{mycolor_sbl}{rgb}{0.92900,0.69400,0.12500}%

\begin{tikzpicture}

\begin{axis}[width=0.78\columnwidth,
height=4.7cm,
tick label style={font=\scriptsize},
scale only axis,
xmin=-16,
xmax=-6,
xlabel style={font=\small\color{white!15!black}},
xlabel={SNR},
ymode=log,
ymin=0.001,
ymax=1,
yminorticks=true,
ylabel style={font=\small\color{white!15!black}},
ylabel={PMD},
axis background/.style={fill=white},
xmajorgrids,
ymajorgrids,
yminorgrids,
legend style={at={(0.03,0.03)}, anchor=south west, legend cell align=left, align=left,  draw=white!15!black,font=\small}
]
\addplot [color=magenta, line width=0.9pt, mark size=2.3pt, mark=diamond, mark options={solid, magenta}]
  table[row sep=crcr]{%
-20	0.926\\
-18	0.861\\
-16	0.789\\
-14	0.655\\
-12	0.561\\
-10	0.47\\
-8	0.397\\
-6	0.275\\
-4	0.321\\
-2	0.398\\
0	0.552\\
2	0.712\\
4	0.943\\
};
\addlegendentry{VAMP}

\addplot [color=mycolor2, line width=0.9pt, mark size=2.0pt, mark=square, mark options={solid, mycolor2}]
  table[row sep=crcr]{%
-20	0.896\\
-18	0.796\\
-16	0.670000000000001\\
-14	0.434\\
-12	0.224\\
-10	0.0759999999999999\\
-8	0.009\\
-6	0.002\\
-4	0\\
-2	0\\
0	0\\
2	0\\
4	0\\
};
\addlegendentry{SOMP}

\addplot [color=mycolor_sbl, line width=0.9pt, mark size=2.5pt, mark=triangle, mark options={solid, mycolor_sbl}]
  table[row sep=crcr]{%
-20	0.796\\
-18	0.634000000000001\\
-16	0.413\\
-14	0.166\\
-12	0.032\\
-10	0.001\\
-8	0\\
-6	0\\
-4	0\\
-2	0\\
0	0\\
2	0\\
4	0\\
};
\addlegendentry{SBL}

\addplot [color=blue, line width=0.9pt, mark size=2.0pt, mark=o, mark options={solid,blue}]
  table[row sep=crcr]{%
-20	0.968\\
-18	0.873\\
-16	0.543\\
-14	0.176\\
-12	0.036\\
-10	0.001\\
-8	0\\
-6	0\\
-4	0\\
-2	0\\
0	0\\
2	0\\
4	0\\
};
\addlegendentry{CL-MP}

\end{axis}
\end{tikzpicture}%
    \caption{PMD as a function of SNR for different algorithms.}
    \label{pmdVsSNRcompare}
\end{figure}
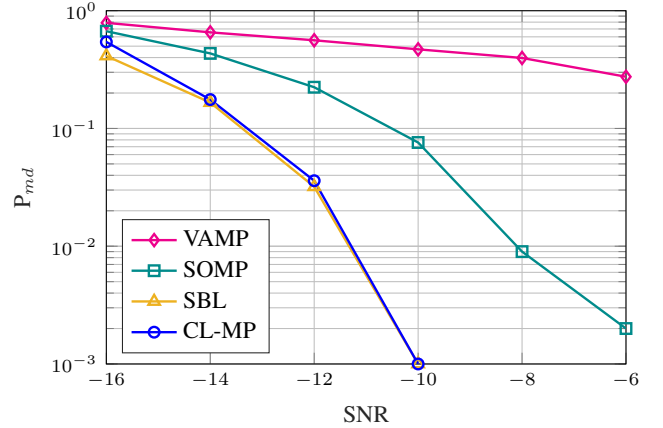
\begin{figure}[!t]
    \centering
%
%
\definecolor{mycolor1}{rgb}{0.54510,0.00000,0.54510}%
\definecolor{mycolor2}{rgb}{0.00000,0.00000,0.50196}%
\definecolor{mycolor3}{rgb}{0.86275,0.07843,0.23529}%
\definecolor{mycolor4}{rgb}{0.18431,0.30980,0.30980}%
\begin{tikzpicture}

\begin{axis}[%
width=0.78\columnwidth,
height=4.8cm,
tick label style={font=\scriptsize} ,
scale only axis,
xmin=-20,
xmax=20,
xlabel style={font=\small\color{white!15!black}},
xlabel={SNR},
ymode=log,
ymin=0.00935999999999991,
ymax=1,
yminorticks=true,
ylabel style={font=\small\color{white!15!black}},
ylabel={PMD},
axis background/.style={fill=white},
xmajorgrids,
ymajorgrids,
yminorgrids,
legend style={legend cell align=left, align=left, draw=white!15!black,at={(1,1)},font=\small}
]
\addplot [color=mycolor1, line width=0.9pt, mark size=2.7pt, mark=o, mark options={solid, mycolor1}]
  table[row sep=crcr]{%
-20	0.986929999999979\\
-16	0.896209999999944\\
-12	0.622360000000003\\
-8	0.2709\\
-4	0.0993600000000081\\
0	0.0784700000000061\\
4	0.0773900000000057\\
8	0.082250000000006\\
12	0.0857300000000055\\
16	0.087930000000006\\
20	0.0883400000000059\\
};
\addlegendentry{RC-FOBM}

\addplot [color=mycolor2, line width=0.9pt, mark size=2.7pt, mark=asterisk, mark options={solid, mycolor2}]
  table[row sep=crcr]{%
-20	0.986939999999979\\
-16	0.893439999999945\\
-12	0.621510000000002\\
-8	0.27151\\
-4	0.0998400000000085\\
0	0.0774700000000063\\
4	0.0785600000000057\\
8	0.083280000000006\\
12	0.0848900000000058\\
16	0.0886000000000058\\
20	0.088280000000006\\
};
\addlegendentry{RC-ROBM}

\addplot [color=mycolor3, line width=0.9pt, mark size=2.7pt, mark=square, mark options={solid, mycolor3}]
  table[row sep=crcr]{%
-20	0.976979999999964\\
-16	0.953129999999942\\
-12	0.882709999999946\\
-8	0.667790000000006\\
-4	0.338650000000013\\
0	0.167509999999998\\
4	0.147759999999999\\
8	0.14901\\
12	0.153129999999998\\
16	0.158699999999995\\
20	0.160449999999996\\
};
\addlegendentry{G-MAT}

\addplot [color=mycolor4, line width=0.9pt, mark size=2.7pt, mark=triangle, mark options={solid, mycolor4}]
  table[row sep=crcr]{%
-20	0.987919999999978\\
-16	0.905439999999945\\
-12	0.480970000000004\\
-8	0.107030000000008\\
-4	0.0119599999999998\\
0	0.00935999999999991\\
4	0.00984999999999989\\
8	0.0101299999999999\\
12	0.0108899999999999\\
16	0.0116499999999998\\
20	0.0115799999999998\\
};
\addlegendentry{B-MAT}

\end{axis}
\end{tikzpicture}%
{}
    \caption{PMD of CL-MP as a function of the SNR.}
    \label{snrPmdPilots}
\end{figure}
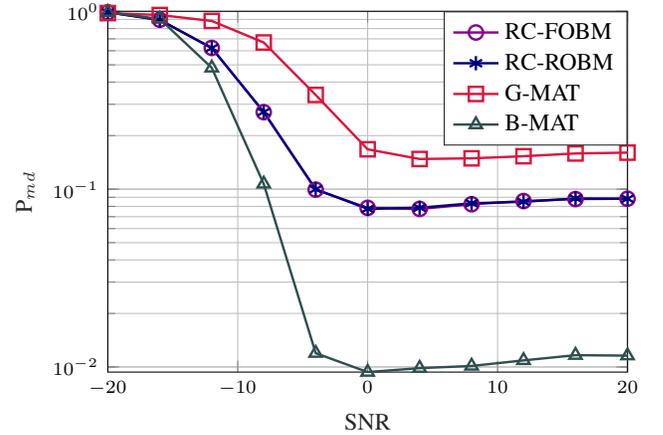

 Fig.~\ref{snrPmdPilots} presents the performance in terms of the probability of miss detection as a function of the SNR and the structure of the pilot sequences used. Similar to the results in Fig.~\ref{pmdVsSNRcompare}, the performance generally improves as the SNR increases. Despite this, the proposed solution CL-MP performs differently for varying sensing matrices. This shows the non-triviality of choosing pilot sequences in MTC \cite{marata2023joint,marata2022joint}. For instance, the performance is superior when using Bernoulli pilot sequences and least when using the Gaussian pilot sequences. This is because pilot sequences such as Gaussian tend to increase the collision probability of the pilot sequences, thus, degrading the performance \cite{marata2023joint,marata2022joint,senel2018grant}. On the other hand, Bernoulli, RC-FOBM, and RC-ROBM have reduced collision probabilities and thus have better performance than the Gaussian pilot sequences. Despite this, the proposed CL-MP performs reasonably well under the different set of pilot sequences. Another interesting result is that the CL-MP performs reasonably well at relatively low SNR values, reaching PMD of=0.01  for BMAT  at SNR of $0$ dB. 
\begin{figure}[!t]
    \centering
%
%
\definecolor{mycolor2}{rgb}{0.00000,0.54510,0.54510}%
\definecolor{mycolor3}{rgb}{0.00000,0.00000,0.50196}%
\definecolor{mycolor_sbl}{rgb}{0.92900,0.69400,0.12500}%
\begin{tikzpicture}
\begin{axis}[%
width=0.78\columnwidth,
height=4.8cm,
scale only axis,
tick label style={font=\scriptsize} ,
xmin=0,
xmax=40,
xlabel style={font=\small\color{white!15!black}},
xlabel={$M$ (number of antennas)},
ymode=log,
ymin=0.001,
ymax=1,
yminorticks=true,
ylabel style={font=\small\color{white!15!black}},
ylabel={PMD},
axis background/.style={fill=white},
xmajorgrids,
ymajorgrids,
yminorgrids,
legend style={at={(0.03,0.03)}, anchor=south west, legend cell align=left, align=left,font=\small,draw=white!15!black}
]
\addplot [color=magenta, line width=0.9pt, mark size=2.3pt, mark=diamond, mark options={solid, magenta}]
  table[row sep=crcr]{%
1	0.859500000000001\\
5	0.67\\
10	0.574\\
15	0.544\\
20	0.5045\\
25	0.4635\\
30	0.481\\
35	0.4725\\
40	0.4685\\
};
\addlegendentry{VAMP}

\addplot [color=mycolor2, line width=0.9pt, mark size=2.3pt, mark=square, mark options={solid, mycolor2}]
  table[row sep=crcr]{%
1	0.907500000000002\\
5	0.6025\\
10	0.366\\
15	0.229\\
20	0.1605\\
25	0.115\\
30	0.0789999999999998\\
35	0.0679999999999999\\
40	0.0589999999999999\\
};
\addlegendentry{SOMP}

\addplot [color=mycolor_sbl, dashed, line width=0.9pt, mark size=2.3pt, mark=triangle, mark options={solid, mycolor_sbl}]
  table[row sep=crcr]{%
1	0.853000000000001\\
5	0.4385\\
10	0.1865\\
15	0.0714999999999999\\
20	0.0155\\
25	0.0055\\
30	0.002\\
35	0.001\\
40	0\\
};
\addlegendentry{SBL}
\addplot [color=blue, line width=0.9pt, mark size=2.3pt, mark=o, mark options={solid, blue}]
  table[row sep=crcr]{%
1	0.873000000000002\\
5	0.455\\
10	0.1815\\
15	0.0689999999999999\\
20	0.017\\
25	0.005\\
30	0.002\\
35	0.001\\
40	0\\
};
\addlegendentry{CL-MP}
\end{axis}
\end{tikzpicture}%
{}
    \caption{PMD as a function of the number of antennas.}
    \label{pmdVsAntennas}
\end{figure}


Fig.~\ref{pmdVsAntennas} shows the performance in terms of PMD as a function of the number of antennas for different algorithms. The general trend shows that the performance of all the algorithms improves with an increase in the number of antennas ($M$). This is mainly because an increase in $M$ imposes more structural sparsity on the matrix $\bm{X}$, thus increasing the accuracy of the estimation. This reveals the advantage of having multiple BS antennas, which leads to a multiple measurement vector (MMV) problem in compressed sensing\cite{senel2018grant}. However, it can be observed that both CL-MP and SBL have equivalent and superior performance compared to VAMP and SOMP, thus, confirming the superiority of our proposed solution. Though both SBL and CL-MP are comparable in terms of performance in PMD, it is worth emphasizing that CL-MP has lower computational complexity than SBL, as shown in Table~\ref{complexAnalysis}, making it a more practical solution in MTC. To substantiate this, we compare the run-times for varying $M$.  

\begin{figure}[!t]
    \centering    
%
%
\definecolor{mycolor_sbl}{rgb}{0.92900,0.69400,0.12500}%
\definecolor{mycolor6}{rgb}{0.00000,0.54510,0.54510}%
\definecolor{mycolor1}{rgb}{0.54510,0.00000,0.54510}%
\definecolor{mycolor2}{rgb}{0.00000,0.00000,0.50196}%
\definecolor{mycolor3}{rgb}{0.86275,0.07843,0.23529}%
\definecolor{mycolor4}{rgb}{0.18431,0.30980,0.30980}%
\begin{tikzpicture}

\begin{axis}[%
width=0.78\columnwidth,
height=4.8cm,
scale only axis,
tick label style={font=\scriptsize} ,
xmin=0,
xmax=40,
xlabel style={font=\color{white!15!black}},
xlabel={$M$ (Number of Antennas)},
ymode=log,
ymin=0.01,
ymax=10,
yminorticks=true,
ylabel style={font=\color{white!15!black}},
ylabel={Runtime (s)},
axis background/.style={fill=white},
xmajorgrids,
ymajorgrids,
yminorgrids,
legend style={at={(0.03,0.03)}, anchor=south west, legend cell align=left, align=left,font=\small,draw=white!15!black}
]
\addplot [color=magenta, line width=0.9pt, mark size=2.3pt, mark=diamond, mark options={solid, magenta}]
  table[row sep=crcr]{%
1	0.440223349\\
5	0.518103036\\
10	0.714324108\\
15	1.137476221\\
20	2.070610011\\
25	3.282435175\\
30	3.890447303\\
35	5.789337936\\
40	7.214156431\\
};
\addlegendentry{VAMP}

\addplot [color=mycolor6, line width=0.9pt, mark size=2.3pt, mark=square, mark options={solid, mycolor6}]
  table[row sep=crcr]{%
1	0.016220192\\
5	0.029197166\\
10	0.03718597\\
15	0.045319514\\
20	0.055049449\\
25	0.061157425\\
30	0.064942267\\
35	0.080294615\\
40	0.086584864\\
};
\addlegendentry{SOMP}

\addplot [color=mycolor6, dashed, line width=0.9pt, mark=triangle, mark options={solid, rotate=270, mycolor6}]
  table[row sep=crcr]{%
1	0.04754175698\\
5	0.07947190042\\
10	0.1046926877\\
15	0.13572120806\\
20	0.15431005664\\
25	0.20000633232\\
30	0.21771580352\\
35	0.23590137764\\
40	0.25971623034\\
};
\addlegendentry{SOMP, $N = 3000$}

\addplot [color=mycolor_sbl, dashed, line width=0.9pt, mark size=2.3pt, mark=triangle, mark options={solid, mycolor_sbl}]
  table[row sep=crcr]{%
1	3.92297428\\
5	4.478195463\\
10	4.435777335\\
15	4.664842833\\
20	4.600409789\\
25	4.704715274\\
30	4.543021942\\
35	4.439072298\\
40	4.35555943\\
};
\addlegendentry{SBL}

\addplot [color=blue, line width=0.9pt, mark=o, mark options={solid, blue}]
  table[row sep=crcr]{%
1	0.00629404588\\
5	0.0160647882\\
10	0.01638674352\\
15	0.01498451582\\
20	0.0144454729\\
25	0.0144625828800001\\
30	0.01448589062\\
35	0.01390095412\\
40	0.0148775464\\
};
\addlegendentry{CB-MP}

\addplot [color=blue, dashed, line width=0.9pt, mark=triangle, mark options={solid, rotate=270, blue}]
  table[row sep=crcr]{%
1	0.1819803739\\
5	0.1755153226\\
10	0.1726382518\\
15	0.17652451936\\
20	0.174582637200001\\
25	0.18188305156\\
30	0.18144085898\\
35	0.17912567478\\
40	0.17784754258\\
};
\addlegendentry{CB-MP, $N= 3000$}

\end{axis}
\end{tikzpicture}%
    \caption{Runtime as a function of the number of antennas.}
    \label{runTimeVsAntenna}
\end{figure}
\par Figure~\ref{runTimeVsAntenna} presents the runtime results of the algorithms for different numbers of antennas. The results show that the proposed CL-MP consistently exhibits lower runtimes compared to VAMP, SBL, and OMP as $M$ increases for $N = 1000$ and $N =3000$. In contrast, algorithms such as VAMP and OMP experience increasing runtimes with the number of antennas.
Notably, although SBL maintains a consistent runtime as $M$ increases, similar to CL-MP, its runtime is more than 100 times that of CL-MP. This is consistent with the fact that CL-MP and SBL have similar cost functions, as discussed in Section~\ref{contributions}. However, despite this similarity, SBL’s significantly higher runtime makes it less favorable for mMTC. Furthermore, CL-MP's runtime for $N = 3000$ is less than the runtime for SBL with $N = 1000$, further highlighting CL-MP' s robustness and consistent computational complexity, making it well-suited for mMTC. 

 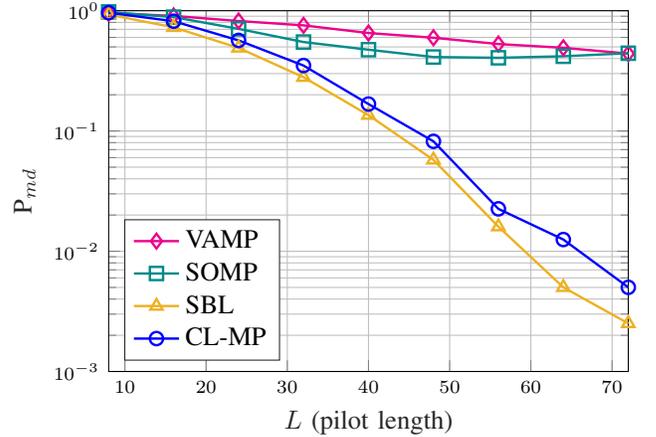
\begin{figure}[!t]
     \centering
%
%
\definecolor{mycolor1}{rgb}{0.54510,0.00000,0.54510}%
\definecolor{mycolor2}{rgb}{0.00000,0.00000,0.50196}%
\definecolor{mycolor3}{rgb}{0.86275,0.07843,0.23529}%
\definecolor{mycolor4}{rgb}{0.18431,0.30980,0.30980}%
\definecolor{mycolor_somp}{rgb}{0.00000,0.54510,0.54510}%
\definecolor{mycolor_sbl}{rgb}{0.92900,0.69400,0.12500}%
\begin{tikzpicture}
\begin{axis}[%
width=0.78\columnwidth,
height=4.8cm,
tick label style={font=\scriptsize} ,
scale only axis,
xmin=8,
xmax=72,
xlabel style={font=\color{white!15!black}},
xlabel={$L$ (pilot length)},
ymode=log,
ymin=0.001,
ymax=1,
yminorticks=true,
ylabel style={font=\color{white!15!black}},
ylabel={PMD},
axis background/.style={fill=white},
xmajorgrids,
ymajorgrids,
yminorgrids,
legend style={at={(0.03,0.03)}, anchor=south west, legend cell align=left, align=left, draw=white!15!black}
]
\addplot [color=magenta, line width=0.9pt, mark=diamond,  mark size = 2.8pt, mark options={solid, magenta}]
  table[row sep=crcr]{%
8	0.957010000000001\\
16	0.905010000000002\\
24	0.82101\\
32	0.75551\\
40	0.65251\\
48	0.59601\\
56	0.52851\\
64	0.49151\\
72	0.440009999999999\\
};
\addlegendentry{VAMP}

\addplot [color=mycolor_somp, line width=0.9pt, mark=square, mark size = 2.5pt, mark options={solid, mycolor_somp}]
  table[row sep=crcr]{%
8	0.976510000000001\\
16	0.884510000000002\\
24	0.706509999999999\\
32	0.54901\\
40	0.47401\\
48	0.41201\\
56	0.40601\\
64	0.41801\\
72	0.44301\\
};
\addlegendentry{SOMP}

\addplot [color=mycolor_sbl, line width=0.9pt, mark=triangle, mark size = 2.7pt,mark options={solid, mycolor_sbl}]
  table[row sep=crcr]{%
8	0.928510000000002\\
16	0.72751\\
24	0.49051\\
32	0.28001\\
40	0.13551\\
48	0.0575099999999999\\
56	0.01601\\
64	0.00501\\
72	0.00251\\
};
\addlegendentry{SBL}

\addplot [color=blue, line width=0.9pt, mark=o, mark size = 2.5pt, mark options={solid, blue}]
  table[row sep=crcr]{%
8	0.962510000000001\\
16	0.818010000000001\\
24	0.56651\\
32	0.35001\\
40	0.16751\\
48	0.0820099999999999\\
56	0.02251\\
64	0.01251\\
72	0.00501\\
};
\addlegendentry{CL-MP}

\end{axis}
\end{tikzpicture}%
     \caption{PMD as a function of the pilot lengths for different algorithms.}
     \label{pmdVsPilot}
 \end{figure}
 
 Fig.~\ref{pmdVsPilot} shows the performance as a function of the pilot lengths ($L$) for different algorithms. The trend reveals that the PMD reduces as the pilot length increases. This is mainly due to better pilot structure as the length increases. Essentially, the signal detection phase is done over a longer period, thus improving the accuracy of the channel estimation for SOMP, VAMP and SBL, as well as the index estimation of the CL-MP. Similar to the results in Fig~\ref{pmdVsAntennas} and Fig.~\ref{pmdVsSNRcompare}, CL-MP and SBL perform better than all the other benchmarks. Consistent with the results of Fig.~\ref{runTimeVsAntenna}, CL-MP has the added advantage of being more scalable.

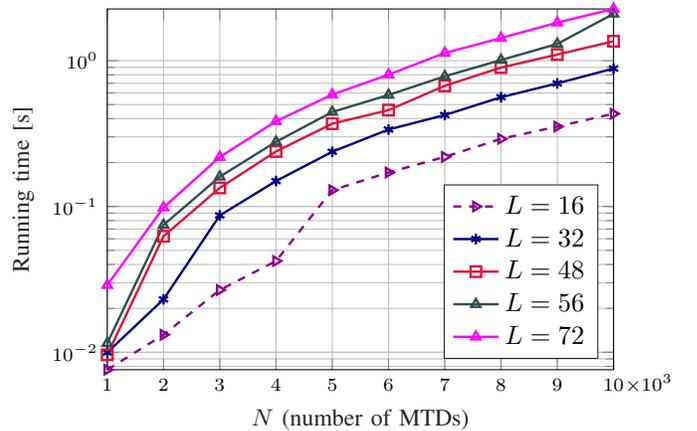
\begin{figure}
    \centering    
%
%
\definecolor{mycolor1}{rgb}{0.54510,0.00000,0.54510}%
\definecolor{mycolor2}{rgb}{0.00000,0.00000,0.50196}%
\definecolor{mycolor3}{rgb}{0.86275,0.07843,0.23529}%
\definecolor{mycolor4}{rgb}{0.18431,0.30980,0.30980}%
\definecolor{mycolor5}{rgb}{1.00000,0.00000,1.00000}%
\begin{tikzpicture}

\begin{axis}[%
width=0.76\columnwidth,
height=4.8cm,
tick label style={font=\scriptsize} ,
xtick={1000,2000,...,10000},
scaled x ticks={real:1000},
xtick scale label code/.code={},
scale only axis,
xmin=1000,
xmax=10000,
xlabel style={font=\small\color{white!15!black}},
xlabel={$N$ (number of MTDs)},
ymode=log,
ymin=0.00758939235000001,
ymax=2.26309622625,
yminorticks=true,
ylabel style={font=\small\color{white!15!black}},
ylabel={Runtime (s)},
axis background/.style={fill=white},
xmajorgrids,
ymajorgrids,
yminorgrids,
legend style={at={(0.97,0.03)}, anchor=south east, legend cell align=left, align=left, draw=white!15!black}
]
\addplot [color=mycolor1, dashed, line width=0.9pt, mark=triangle, mark options={solid, rotate=270, mycolor1}]
  table[row sep=crcr]{%
1000	0.00758939235000001\\
2000	0.0131768608\\
3000	0.0267357934\\
4000	0.04225318775\\
5000	0.1294140952\\
6000	0.1710331739\\
7000	0.21877810885\\
8000	0.29099267475\\
9000	0.3530463022\\
10000	0.432806050649999\\
};
\addlegendentry{$L = 16$}

\addplot [color=mycolor2, line width=0.9pt, mark=asterisk, mark options={solid, mycolor2}]
  table[row sep=crcr]{%
1000	0.00996678094999998\\
2000	0.0230579189499999\\
3000	0.086790021\\
4000	0.1494244102\\
5000	0.23724667085\\
6000	0.3371062426\\
7000	0.423582552099999\\
8000	0.561270168149999\\
9000	0.6985181394\\
10000	0.8829378312\\
};
\addlegendentry{$L = 32$}

\addplot [color=mycolor3, line width=0.9pt, mark=square, mark options={solid, mycolor3}]
  table[row sep=crcr]{%
1000	0.00960332235000001\\
2000	0.06249717125\\
3000	0.1339943726\\
4000	0.2383884279\\
5000	0.369731887500001\\
6000	0.45783637355\\
7000	0.6723114794\\
8000	0.896601474100001\\
9000	1.10107499945\\
10000	1.35838783455\\
};
\addlegendentry{$L = 48$}

\addplot [color=mycolor4, line width=0.9pt, mark=triangle, mark options={solid, mycolor4}]
  table[row sep=crcr]{%
1000	0.01153165455\\
2000	0.0746647405000001\\
3000	0.1599248007\\
4000	0.27766429225\\
5000	0.444657437399999\\
6000	0.58144406025\\
7000	0.78133738625\\
8000	1.0094133153\\
9000	1.30231390935\\
10000	2.09626752\\
};
\addlegendentry{$L =  56$}

\addplot [color=mycolor5, line width=0.9pt, mark=triangle, mark options={solid, mycolor5}]
  table[row sep=crcr]{%
1000	0.02886235875\\
2000	0.0982442171500002\\
3000	0.21813909545\\
4000	0.383903643299999\\
5000	0.5847110155\\
6000	0.801427598449998\\
7000	1.12770693435\\
8000	1.4294977322\\
9000	1.822223672\\
10000	2.26309622625\\
};
\addlegendentry{$L = 72$}

\end{axis}
\node at (7.2, -.185) (a) {\scriptsize $\times 10^3$};
\end{tikzpicture}%
    \caption{Runtime of CL-MP as a function of the number $N$ of the MTDs in the network.}
    \label{runTimeVsUsers}
\end{figure}
\par To conclude the results, in Fig.~\ref{runTimeVsUsers} we present the runtime of CL-MP as a function of the number of MTDs. In general, the runtime increases with the number of MTDs, which is expected because the size of $\boldsymbol{\gamma}$ increases. 
Despite this, the CL-MP has lower runtimes for a large number of users. For instance, it takes approximately $1$ second for CL-MP to recover and perform AUD from the signal received from $N= 6000$ users when the pilot length is  $L = 72$. This translates to better runtime than SBL, which is shown to require more than $5$ seconds to process signals from $N = 1000$ users in Fig.~\ref{runTimeVsAntenna}. To further support the results of Fig.~\ref{runTimeVsUsers}, in Fig.~\ref{pmdVsNumberOfMTDs}, we present the results of PMD for varying sizes of the networks. In general, the proposed CL-MP algorithm is effective in solving AUD problems, with very low PMD even in very large MTC networks, rendering our proposal a practical solution for the rapidly growing MTC networks.   
\begin{figure}[!t]
     \centering     
%
%
\definecolor{mycolor1}{rgb}{0.54510,0.00000,0.54510}%
\definecolor{mycolor2}{rgb}{0.00000,0.00000,0.50196}%
\definecolor{mycolor3}{rgb}{0.86275,0.07843,0.23529}%
\definecolor{mycolor4}{rgb}{0.18431,0.30980,0.30980}%
\definecolor{mycolor5}{rgb}{1.00000,0.00000,1.00000}%
\begin{tikzpicture}

\begin{axis}[%
width=0.78\columnwidth,
height=4.8cm,
tick label style={font=\scriptsize},
xtick={1000,2000,...,10000},
scaled x ticks={real:1000},
xtick scale label code/.code={},
scale only axis,
xmin=1000,
xmax=10000,
xlabel style={font=\small\color{white!15!black}},
xlabel={$N$ (number of MTDs)},
ymin=0,
ymax=1,
ylabel style={font=\small\color{white!15!black}},
ylabel={PMD},
axis background/.style={fill=white},
xmajorgrids,
ymajorgrids,
legend style={at={(1,0.0)}, anchor=south east, legend cell align=left, align=left, opacity=0.85,font=\small,draw=white!15!black}
]
\addplot [color=mycolor1, dashed, line width=0.9pt, mark=triangle, mark options={solid, rotate=270, mycolor1}]
  table[row sep=crcr]{%
1000	0.00969999999999999\\
2000	0.634875\\
3000	0.819283333333324\\
4000	0.877749999999999\\
5000	0.909510000000019\\
6000	0.928758333333355\\
7000	0.94099999999999\\
8000	0.948556250000003\\
9000	0.955938888888888\\
10000	0.960765000000023\\
};
\addlegendentry{$L = 16$}

\addplot [color=mycolor2, line width=0.9pt, mark=asterisk, mark options={solid, mycolor2}]
  table[row sep=crcr]{%
1000	0\\
2000	0.00025\\
3000	0.0998333333333319\\
4000	0.547399999999999\\
5000	0.686170000000002\\
6000	0.76305833333333\\
7000	0.809528571428569\\
8000	0.841481249999999\\
9000	0.866222222222218\\
10000	0.884225000000008\\
};
\addlegendentry{$L = 32$}

\addplot [color=mycolor3, line width=0.9pt, mark=square, mark options={solid, mycolor3}]
  table[row sep=crcr]{%
1000	0\\
2000	0\\
3000	1.66666666666667e-05\\
4000	0.00408750000000002\\
5000	0.25475\\
6000	0.514708333333334\\
7000	0.632128571428569\\
8000	0.70296875\\
9000	0.748922222222222\\
10000	0.786174999999987\\
};
\addlegendentry{$L = 48$}

\addplot [color=mycolor4, line width=0.9pt, mark=triangle, mark options={solid, mycolor4}]
  table[row sep=crcr]{%
1000	0\\
2000	0\\
3000	0\\
4000	0.0001\\
5000	0.00998999999999987\\
6000	0.306775\\
7000	0.511857142857143\\
8000	0.613974999999999\\
9000	0.681405555555556\\
10000	0.728175000000005\\
};
\addlegendentry{$L =  56$}

\addplot [color=mycolor5, line width=0.9pt, mark=triangle, mark options={solid, mycolor5}]
  table[row sep=crcr]{%
1000	0\\
2000	0\\
3000	0\\
4000	0\\
5000	0\\
6000	0.00108333333333333\\
7000	0.053878571428572\\
8000	0.35775625\\
9000	0.504183333333333\\
10000	0.593449999999997\\
};
\addlegendentry{$L = 72$}
\end{axis}
\node at (7.5, -.185) (a) {\scriptsize $\times 10^3$};
\end{tikzpicture}%
     \caption{PMD as a function of the number  $N$ of the MTDs in the network.}
     \label{pmdVsNumberOfMTDs}
 \end{figure}
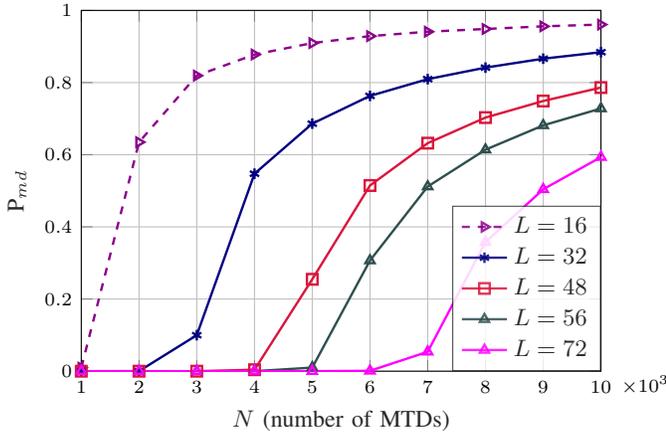

\section{Conclusions and Future Directions}
\label{conclusionAndFuture}
This paper introduces a novel framework for estimating the active user indices in MTC. The proposed technique uses the covariance of the received signal to formulate an optimization problem for the sparse effective power levels in MTC and solves it using a greedy approach. The results demonstrate that the proposed framework offers highly accurate device activity detection with very low runtimes. However, while efficient, the proposed solution does not estimate the activation probability. Therefore, the results of this work can be extended to incorporate activation probability estimation as future research direction. Moreover, considering that some IoT networks will
eventually be served by Low Earth Orbit (LEO) satellites, which are power constrained, one
potential extension of this work would be the integration of
CL-MP in satellite receivers to exploit its computational efficiency. In this case, the CL-MP can be adapted to incorporate the line of sight conditions, which can potentially minimize interference and thus yield better recovery.

\appendix 
\subsection{Minimizer of  the conditional negative LLF } \label{App:A}

First we note the following generic derivative rules: 
\begin{align}
\frac{\partial \mathbf \M^{-1}}{\partial \gamma }  &= - \M^{-1} \frac{\partial  \M}{\partial \gamma }   \M^{-1}  \label{eq:app:der_sigmainv} \\
\frac{\partial  \log | \M | }{\partial \gamma }&= \tr \left( \M^{-1}  \frac{\partial  \M}{\partial \gamma }  \right)
\end{align} 
Since $\M = \M_{\setminus i } + \gamma \a_i \a^\hop_i$, we get $  \frac{\partial  \M}{\partial \gamma }  = \a_i \a^\hop_i$, and thus 
\begin{align*}
\frac{\partial\tr (\mathbf \M^{-1} \hat \M )}{\partial \gamma }  &= - \a_i^\hop \M^{-1} \hat \M \M^{-1} \a_i,  \ \ \frac{\partial \log | \M |}{\partial \gamma } &= \a_i^\hop \M^{-1} \a_i. 
\end{align*} 
Hence the derivative of $\ell_i(\gamma  \mid \M_{\setminus i })= \tr(  \M^{-1} \SCM )  +  \log | \M |  $ is 
\begin{align}
\ell_i'(\gamma  \mid \M_{\setminus i }) &=  \a_i^\hop \M^{-1} \a_i -  \a_i^\hop \M^{-1} \hat \M \M^{-1} \a_i .   \label{app:apueq1}
 \end{align} 
Next note that 
\beq \label{app:apueq2} 
\M^{-1}\a_i = (\M_{\setminus i } + \gamma \a_i \a^\hop_i)^{-1} \a_i =  \frac{ \M_{\setminus i }^{-1} \a_i}{1+ \gamma \a_i^\hop \M_{\setminus i }^{-1} \a_i}
\eeq 
which follows by multiplying both sides of Sherman-Morrison formula \eqref{eq:Minv_update} by $\a$ from the right, and simplifying the resulting expression. Applying \eqref{app:apueq2} once to the first term and twice to the second term in \eqref{app:apueq1} yields:
\beq
\ell_i'(\gamma \mid \M_{\setminus i })=\frac{\a_i^\hop \M_{\setminus i } \a_i}{1+  \gamma \a_i^\hop \M_{\setminus i } \a_i}  - \frac{\a_i^\hop  \M_{\setminus i }^{-1} \hat \M   \M_{\setminus i }^{-1} \a_i}{(1+ \gamma \a_i^\hop \M_{\setminus i } \a_i)^2} \label{app:apueq1b} .
\eeq
Setting the derivative to zero, $\ell_i(\tilde \gamma  \mid \M_{\setminus i })=0$, shows that the minimizer  must verify:
\beq \label{eq:app:root}
\frac{\a_i^\hop \M_{\setminus i } \a_i}{1+  \tilde \gamma \a_i^\hop \M_{\setminus i } \a_i}  - \frac{\a_i^\hop  \M_{\setminus i }^{-1} \hat \M   \M_{\setminus i }^{-1} \a_i}{(1+ \tilde \gamma \a_i^\hop \M_{\setminus i } \a_i)^2}  = 0 . 
\eeq 
Multiplying both sides by $(1+ \tilde \gamma \a_i^\hop \M_{\setminus i } \a_i)^2$ and solving for $\tilde \gamma$ shows that \eqref{eq:app:root} has a  single root:
\beq \label{eq:app:solution} 
\tilde \gamma =  \frac{\a_i^\hop \M_{\setminus i}^{-1}  \S  \M_{\setminus i}^{-1} \a_i - \a_i^\hop \M_{\setminus i}^{-1} \a_i }{(\a_i^\hop \M_{\setminus i}^{-1} \a_i)^2 }.
\eeq 
Taking derivative of  $\ell_i'(\gamma  \mid \M_{\setminus i })$  in \eqref{app:apueq1} and using the derivative rule \eqref{eq:app:der_sigmainv} gives the following expression for the 2nd derivative:
\beq \label{eq:app:lprimeprime}
\ell^{\prime\prime}_i(\gamma  \mid \M_{\setminus i })=  - (\a_i^\hop \M^{-1} \a_i)^2 + 2 (\a_i^\hop \M^{-1} \a_i)  \a_i^\hop \M^{-1} \hat \M \M^{-1} \a_i .
\eeq 
Applying \eqref{app:apueq2} in \eqref{eq:app:lprimeprime}  allows to write it in the form
\begin{align*}
\ell^{\prime\prime}_i(\gamma \mid \M_{\setminus i })  
&=   - \frac{(\a_i^\hop \M^{-1}_{\setminus i} \a_i)^2}{(1+ \gamma \a_i^\hop \M_{\setminus i } \a_i)^2} \\
&\qquad +  2 \frac{(\a_i^\hop \M^{-1}_{\setminus i}  \a_i)  \a_i^\hop \M^{-1}_{\setminus i} \hat{\M} \M^{-1}_{\setminus i}  \a_i }{(1+ \gamma \a_i^\hop \M_{\setminus i } \a_i)^3} .
\end{align*}
Hence  value of $\ell^{\prime\prime}_i(\gamma) $ at  the root $\tilde \gamma$ in \eqref{eq:app:solution} is 
\[
\ell^{\prime\prime}_i(\tilde \gamma \mid \M_{\setminus i })   = \left[ \frac{  \a_i^\hop \M^{-1}_{\setminus i} \a_i }{ 1+ \hat \gamma \a_i^\hop \M_{\setminus i } \a_i} \right]^2
\]
Since $\ell^{\prime\prime}_i(\tilde \gamma\mid \M_{\setminus i })  > 0$ for all $\tilde \gamma \neq 0$,   the root is the unique minimizer of $\ell_i(\gamma\mid \M_{\setminus i })$. 

If  the minimizer is negative, i.e., $\tilde \gamma <0$, then since the minimizer is unique, it holds that $\ell_i(0\mid \M_{\setminus i }) < \ell_i(\gamma\mid \M_{\setminus i })$  $\forall \gamma \geq 0$.  This reveals that  the minimizer under the non-negativity constraint ($\gamma \geq 0$) is 
$\hat \gamma = \max(\tilde \gamma,0)$, which is the equation expressed in  \eqref{eq:gamma_i_star}.

\ifCLASSOPTIONcaptionsoff
  \newpage
\fi
\bibliographystyle{IEEEtran} 

\end{document}